\documentclass[aps,prx,a4paper,superscriptaddress,a4paper,twocolumn,showkeys,longbibliography]{revtex4-2}

\usepackage[section]{placeins}

\usepackage{lineno}
\usepackage{graphicx}  % needed for figures
\usepackage{dcolumn}   % needed for some tables
\usepackage{bm}        % for math
\usepackage{amssymb}   % for math
\usepackage{amsmath,stmaryrd}
\usepackage{blkarray, multirow, graphicx, diagbox, color, xcolor, colortbl}
\usepackage{bbm, bbold}
\usepackage{glossaries}
\usepackage{hyphenat}
\usepackage{ifthen}
\usepackage[colorlinks, linkcolor = blue, citecolor = blue, filecolor = black, urlcolor = blue]{hyperref}
\usepackage{xkeyval}
\usepackage{moreverb}
\usepackage{rotating}
\usepackage{wrapfig}
\usepackage{slashbox}
\usepackage{xspace}
\usepackage{nicefrac}
\usepackage[]{units}
\usepackage{physics}
\usepackage{booktabs}
\usepackage{braket}
\usepackage[inline]{enumitem}
\usepackage{tabto}
\usepackage{listings}
\usepackage{xstring}
\def\ReplaceStr#1{%
	\IfSubStr{#1}{p}{%
		\StrSubstitute{#1}{p}{.}}{#1}}

\usepackage[caption=false]{subfig}
\usepackage[capitalise]{cleveref} %
\usepackage{chemformula}
\usepackage{algorithm}

\newacronym{CDW}{CDW}{charge\hyp density wave}
\newacronym{1D}{1D}{one\hyp dimensional}
\newacronym{iTEBD}{iTEBD}{infinite time-evolving block decimation}
\newacronym{TDVP}{TDVP}{time-dependent variational principle}
\newacronym{1TDVP}{1TDVP}{single-site time-dependent variational principle}
\newacronym{2TDVP}{2TDVP}{two-site time-dependent variational principle}
\newacronym{GSE-TDVP}{GSE-TDVP}{global subspace expansion time-dependent variational principle}
\newacronym{LSE-TDVP}{LSE-TDVP}{local subspace expansion time-dependent variational principle}
\newacronym{PP-TDVP}{PP-TDVP}{projected purified time-dependent variational principle}
\newacronym{PP}{PP}{projected purification}
\newacronym{PP-MPS}{PP-MPS}{projected\hyp purified matrix\hyp product states}
\newacronym{GSE}{GSE}{global subspace expansion}
\newacronym{MPS}{MPS}{matrix\hyp product state}
\newacronym{1RDM}{1RDM}{single\hyp site reduced density\hyp matrix}
\newacronym{DMRG}{DMRG}{density\hyp matrix renormalization group}
\newacronym{PST}{PST}{phonon state tomography}
\newacronym{SVD}{SVD}{singular value decomposition}
\newacronym{OBC}{OBC}{open boundary conditions}
\newacronym{PBC}{PBC}{periodic boundary conditions}
\newacronym{TDS}{TDS}{thermal diffuse scattering}
\newacronym{PSS}{PSS}{perfect sampling scheme}
\newacronym{SA-MR-PDF}{SA-MR-PDF}{sample\hyp averaged, momentum\hyp resolved phonon distribution function}
\newacronym{MPO}{MPO}{matrix\hyp product operator}

\usepackage[normalem]{ulem}
\graphicspath{{figures_maintext/}}

\begin{document}

\def\thetitle{Ultrafast electronic coherence from slow phonons}

\title{\thetitle}

\author{M.~Moroder}
\affiliation{School of Physics, Trinity College Dublin, College Green, Dublin 2, D02K8N4, Ireland}

\author{S.~Paeckel}
\affiliation{Department of Physics, Arnold Sommerfeld Center for Theoretical Physics (ASC), Munich Center for Quantum Science and Technology (MCQST), Ludwig-Maximilians-Universit\"{a}t M\"{u}nchen, 80333 M\"{u}nchen, Germany}

\author{M.~Mitrano}
\affiliation{Department of Physics, Harvard University, Cambridge, Massachusetts 02138, USA}

\author{J.~Sous}\email{Author to whom correspondence should be addressed: john.sous@yale.edu.}
\affiliation{Department of Applied Physics, Yale University, New Haven, Connecticut 06511, USA}
\affiliation{Energy Sciences Institute, Yale University, West Haven, Connecticut 06516, USA}

\maketitle

\noindent
\textbf{
Light offers a route to engineer new phases of matter far from equilibrium, including transient states suggestive of superconducting, charge-ordered, and excitonic ordering behavior. Yet it remains unclear how optical excitation can dynamically produce long-range phase coherence—a defining feature of true order such as superconductivity—rather than merely enhancing local pairing. Here we show that impulsively driven low-frequency phonons enhance long-range electronic correlations in a low-dimensional metal. Through numerically exact simulations, we demonstrate that slow phonons suppress dynamical disorder, enabling buildup of coherence and enhancement of charge (and pairing) orders. These findings provide direct evidence that light can mediate enhancement of long-range order and suggest that future experimental strategies—such as the design of selective excitations of narrow phonon distributions to limit dephasing—may offer viable routes to design and stabilize transient superconducting states.
}

The prospect of light-induced phases of matter, such as superconductivity~\cite{Mitrano2016}, charge ordering~\cite{Kogar2020} and excitonic insulation~\cite{Baldini2023} has catalyzed intense interest across condensed matter physics. Although no fundamental principles forbid a symmetry-broken state from emerging far from equilibrium, the mechanisms by which ultrafast optical excitation might drive the formation of phase-coherent order remain unresolved. A key puzzle is how injecting energy—typically expected to thermalize and suppress ordering—can instead lead to enhanced correlations and macroscopic coherence. Another central question is whether light-induced coherence genuinely exhibits long-range order akin to its equilibrium counterpart, or represents a distinct, fluctuating state.

In the specific case of superconductivity, experimental reports have revealed transient signatures in optics, critical currents and Meissner magnetic field expulsion evocative of superconductivity at temperatures well above the equilibrium critical temperature $T_c$, achieved via selective phonon excitation in materials such as cuprates~\cite{Fausti2011, Liu2020}, fullerides~\cite{Mitrano2016,Budden2021} and the $\kappa$ organic salts~\cite{Buzzi2020}. These studies indicate that coherent optical phonon driving provide access to hidden electronic phases unreachable under equilibrium conditions. The diverse nature of these states has been attributed to various mechanisms, including the amplification of superconducting pairing~\cite{Knap2016, Kennes2017}, effective cooling of quasiparticles~\cite{Nava2018}, and suppression of competing orders~\cite{Sentefcompeting2017}. Nonetheless, unambiguous evidence for long-range superconducting coherence remains elusive, and the dynamical mechanism behind the apparent ordering remains under debate~\cite{Chiriaco2018,Dai2021}.

\begin{figure*}
\centering\includegraphics[width=2\columnwidth]{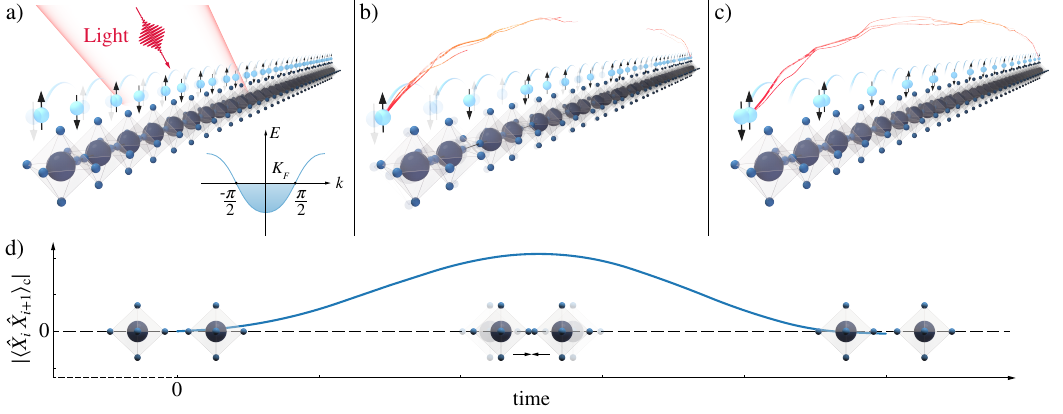}
\caption{\textbf{Light-enhanced long-range electronic correlations.} Schematic illustration showing the effect of an impulsive optical pulse that excites IR-active phonons in a half-filled metal at initial time (panel a). This excitation dynamically induces enhanced long-range correlations at later times (panels b and c), even when the driving is insufficient to generate large local double occupancy. The resulting state exhibits strong tendencies toward charge ordering with a staggered density profile and an associated propensity for lattice dimerization, as evidenced by the connected phonon-phonon correlator $\langle \hat{X}_i \hat{X}_{i+1} \rangle_\mathrm{c}$ (panel d), see Supplementary Materials.
}
\label{fig:fig1}
\end{figure*}

Capturing the non-equilibrium dynamics of light-driven superconductivity remains a formidable challenge for theory. This requires tracking the collective evolution of a strongly correlated many-body system under ultrafast, nonlinear excitation, where both electrons and phonons are driven far from equilibrium. Most theoretical treatments focus on enhanced pairing amplitudes, typically quantified via double occupancy or related local observables, following pump-induced modification of electron-phonon interactions~\cite{Knap2016,Sentef2016,Babadi2017,Coulthard2017,Kennes2017,Murakami2017,Sentef2017,eckhardt2024,chattopadhyay2025}. However, these local measures fall short of diagnosing true superconductivity, which requires long-range phase coherence—a collective reorganization of the entire many-body state. Methodologically, the problem is compounded by severe limitations. Floquet-based analytical methods fail for broadband, non-periodic excitations, and strong pulses  excite large phonon populations, limiting most numerical techniques to small systems. Matrix product state (MPS) simulations enable access to larger systems and have shown that light can dynamically generate disorder that suppresses long-range electronic coherence, presenting a fundamental obstacle to light-induced superconductivity~\cite{Sous2021}. Yet, this result was restricted to fast-phonon regimes, where the phonon frequency exceeds the electronic hopping, leaving open the question of whether slow phonons relevant to real materials might enable a different outcome.  Nonetheless, this highlights emergent dynamical disorder as a novel feature of light-induced non-equilibrium dynamics, calling for methods to identify conditions under which disorder can be suppressed or avoided altogether—if at all possible—reframing the pursuit of light-induced order as a competition between induced correlations and dynamical dephasing. 

Importantly, efforts to induce or enhance superconductivity via ultrafast light pulses in experiment have typically relied on the selective excitation of lattice vibrations with energies smaller than the electronic energy scale. In layered cuprates such as La$_{1.8-x}$Eu$_{0.2}$Sr$_x$CuO$_4$~\cite{Fausti2011} and YBa$_2$Cu$_3$O$_x$~\cite{Hu2014optically,Kaiser2014optically}, key infrared (IR)-active modes include in-plane bond-stretching and apical oxygen phonons in the 70--80~meV range. In fullerides such as K$_3$C$_{60}$~\cite{Mitrano2016,Budden2021} and organic superconductors like $\kappa$-(BEDT-TTF)$_2$Cu[N(CN)$_2$]Br~\cite{Buzzi2020photomolecular}, the relevant excitations are higher-frequency molecular vibrations around 170--180~meV, yet still smaller than the electronic energy scale. Thus, a rigorous assessment of the competition between dynamical dephasing and enhanced correlations requires numerically exact methods capable of capturing both the microscopic dynamics and macroscopic coherence to reveal the conditions under which light can truly stabilize long-range ordered phases in real materials.

Here, we present a numerically exact analysis showing that optical excitation of slow phonons enhances long-range coherence in an optically driven metal (Figure~\ref{fig:fig1}). We show that driven slow phonons diminish disorder, enabling enhancement of  charge correlations—even when local pairing remains only modestly enhanced. This highlights a crucial and previously underappreciated role of the phonon energy scale. Although both the effective attraction and the disorder induced by the pump scale with the number of excited phonons, we show that the disorder component is significantly suppressed at low phonon frequencies. We rationalize this behavior by showing that the pumped phonons generate a distribution of disorder potentials whose width is governed by the phonon frequency. Slow phonons produce narrower phonon number distributions and therefore weaker disorder. In this regime, electrons retain sufficient mobility to phase-lock, despite moderate fluctuations and weak attraction.  

This progress is enabled by an MPS method based on projected purification (PP)~\cite{Koehler2021}, tailored to simulate the full quantum dynamics of nonlinearly coupled electron-phonon systems driven far from equilibrium, with fully converged phonon sectors across the entire range of phonon energy scales and for sufficiently large systems sizes, enabling us to resolve the spatial structure of emergent correlations in the most experimentally germane limit of slow phonons. 
Importantly, our simulations constitute the first numerically exact demonstration of light-induced enhancement of long-range phase coherence in a driven material that would otherwise remain orderless in a ‘vegetative’ metallic state. This represents a controlled, non-perturbative confirmation that serves to suggest that enhancement of true long-range coherence is feasible and is likely to stabilize ordered phases of matter such as superconductivity in higher dimensions upon optical driving under specific dynamical conditions.

\begin{figure*}
\centering
\includegraphics[width=1.7\columnwidth]{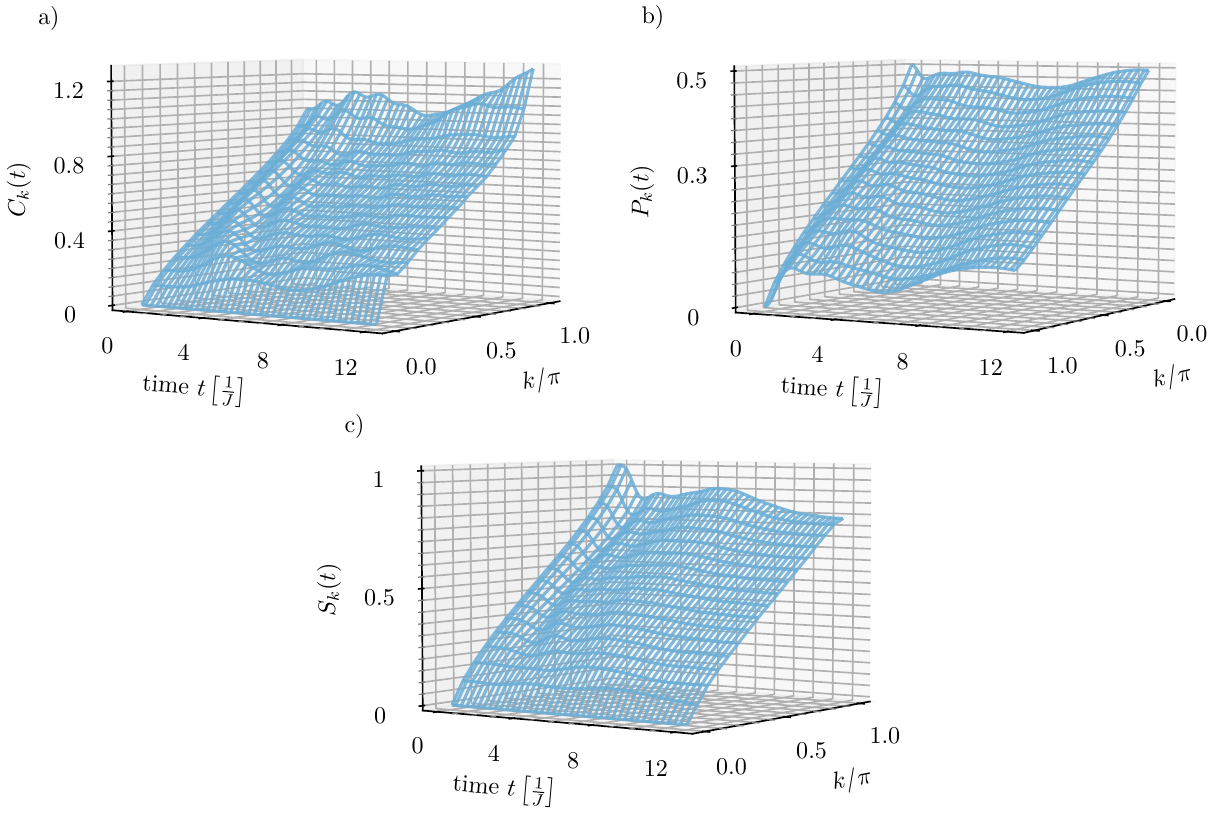}
\caption{\textbf{Pump-induced dynamics of momentum-resolved electron correlations.} We study the evolution with time $t$ of the Fourier transform $\mathcal{F}$ of the equal-time charge correlator $C_k(t) = \mathcal{F}[\langle \hat{n}_i \hat{n}_{i+r} \rangle_\mathrm{c}]$ (panel a),  pairing correlator $P_k(t) = \mathcal{F}[\langle \hat{c}^\dagger_{i,\uparrow} \hat{c}^\dagger_{i,\downarrow} \hat{c}_{i+r,\downarrow} \hat{c}_{i+r,\uparrow} \rangle]$ (panel b) and spin correlator $S_k(t) = \mathcal{F}[\langle (\hat{n}_{i,\uparrow} - \hat{n}_{i,\downarrow})(\hat{n}_{i+r,\uparrow} - \hat{n}_{i+r,\downarrow}) \rangle]$ (panel c) for $g_q / \omega = 0.7/\pi \approx 0.22$  and $\omega = \pi/10$ corresponding to the slow phonon regime.  Here, $\langle \cdot \rangle_\mathrm{c}$ denotes the connected correlator. These results were obtained for systems with $L=32$ sites. Note the $k$-axis of the $P_k(t)$ plot in panel b has been inverted for better visibility. Notice that $C_0(t)$ and $S_0(t)$ correspond to the charge and spin density respectively, and thus are conserved in the dynamics. A rapid increase of approximately 20\% in $C_\pi(t)$, accompanied by a modest rise in $P_k(t)$ especially for small $k$ and a comparably strong suppression of $S_\pi(t)$ indicate an enhanced tendency toward long-range staggered charge density ordering of the initial  $t=0$ metallic state. 
}
\label{fig:fig2}
\end{figure*}

\bigskip

\noindent\textbf{Model of impulsively driven low-dimensional metal}

\noindent In a centrosymmetric crystal, an optically excited dipole-active phonon must couple nonlinearly to the electronic charge density. To model this physics in an impulsively excited low-dimensional metal, we consider a one-dimensional (1D) system of electrons coupled nonlinearly to phonons in a metallic state subject to the application of a short-duration optical pulse at initial time. To leading order, the interactions between electrons and phonons are quadratic,  governed by the Hamiltonian
\begin{equation}
\hat{H} = \hat{H}_{\rm e} + \hat{H}_{\rm ph} + \hat{H}_{\rm e\mbox{-}ph},
\label{Eq:QephHam}
\end{equation}
where $\hat{H}_{\rm e} = -J \sum_{i,\sigma} \hat{c}^\dagger_{i,\sigma} \hat{c}_{i+1,\sigma} + \mathrm{H.c.}$ describes spinful tight-binding electrons with creation (annihilation) operators $\hat{c}^\dagger_{i,\sigma}$ ($\hat{c}_{i,\sigma}$) at site $i$ and spin $\sigma \in \{\uparrow,\downarrow\}$. Phonons are modeled as Einstein oscillators with frequency $\omega$ ($\hbar=1$) via $\hat{H}_{\rm ph} = \omega \sum_i \left( \hat{b}^\dagger_i \hat{b}_i + \tfrac{1}{2} \right)$, where $\hat{b}^\dagger_i$ ($\hat{b}_i$) creates (annihilates) a boson on site $i$. Electron-phonon coupling enters through the term,
\begin{equation}
\hat{H}_{\rm e\mbox{-}ph} = g_q \sum_i (\hat{n}_i - 1)(\hat{b}^\dagger_i + \hat{b}_i)^2,
\end{equation}
where $\hat{n}_i = \sum_\sigma \hat{c}^\dagger_{i,\sigma} \hat{c}_{i,\sigma}$. This model has been proposed as a minimal model for optically driven electron-phonon dynamics~\cite{Kennes2017}. In the atomic limit, the coupling renormalizes the oscillator stiffness as $K \rightarrow K [1 + 4(g_q/\omega)(n - 1)]$ ($n$ is the electron density), which imposes the stability constraint $-\omega/4 < g_q  < \omega/4 $~\cite{Kennes2017,Sous2021}. Within this range, the ground state at half filling $\langle \hat{n}_i \rangle = 1$ remains metallic and nearly indistinguishable from a `free' Fermi sea $\ket{\mathrm{FS}}$, yet the system exhibits pronounced non-equilibrium correlation dynamics absent in linear models~\cite{Sous2021}.

Here, we model contemporary pump-probe experiments in which long-wavelength ultrafast laser pulses act as spatially uniform displacement fields that coherently excite dipole-active phonon modes across the entire crystal lattice. This is implemented by applying a global displacement operator to the initial state, $\hat{\mathcal{D}}(\alpha) = \prod_i \hat{D}_i(\alpha), \quad \text{with} \quad \hat{D}_i(\alpha) = \exp\left[\alpha \hat{b}^\dagger_i - \alpha^* \hat{b}_i\right]$, where $|\alpha|^2$ encodes the average number of phonons excited per site and serves as a proxy for the pump fluence. We initialize the system in a metallic state at zero temperature obtained by finding the ground state of the Hamiltonian, and apply the displacement operator as a quench, initiating non-equilibrium dynamics. The zero temperature framework is appropriate to describe systems in which $\hbar \omega \ll k_B T$ such that the initial phonon state is effectively the vacuum, see Refs.~\cite{Mitrano2016, Buzzi2020}. We track the time evolution of observables and correlation functions to characterize the ensuing dynamics, evolving the system using large-scale MPS simulations (see \textbf{Methods} for details).

\begin{figure*}[hbt]
\centering\includegraphics[width=1.9\columnwidth]{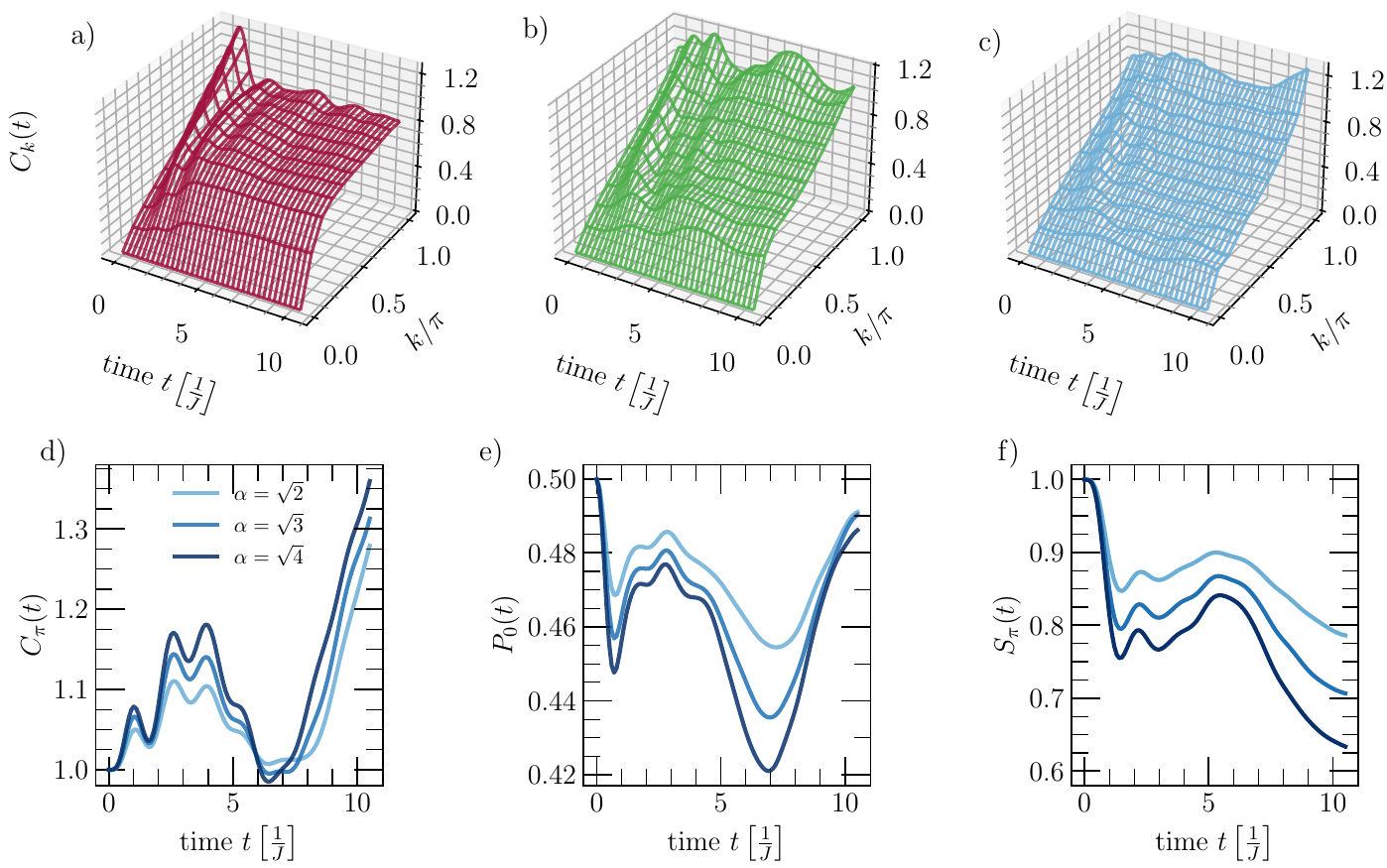}
\caption{\textbf{Phenomenology of light-enhanced long-range electronic correlation.} In the upper row, we examine the dynamics of the momentum-resolved charge correlation $C_k(t)$ as the phonon frequency decreases (from panel a to panel c), with parameters $\omega = (\pi/2)J$, $(\pi/5)J$, and $(\pi/10)J$ in panels a, b, and c, respectively, and corresponding couplings $g_q = 0.35J$, $0.14J$, and $0.07J$, such that the ratio $g_q/\omega$ remains fixed. In the lower row, we study the influence of the pump fluence $|\alpha|^2$ on the staggered charge correlation $C_\pi(t)$(panel d), uniform pairing correction $P_0(t)$ (panel e) and staggered spin correlation $S_\pi(t)$ (panel f) for the smallest phonon frequency we simulate, $\omega = \pi/10 J$, and using $g_q = 0.07J$. These results were obtained for systems with size $L=20$ sites. Fast phonons lead to dynamic disorder, which manifests as a flattening of momentum-space charge correlations (panel~a), whereas slow phonons give rise to coherence at momentum $\pi$, indicating the development of long-range ordering tendencies (panel~c). These enhanced correlations scale with $\alpha$ (panel~d), in contrast to $P_0(t)$, which exhibits increasing fluctuations with larger $\alpha$ (panel~e), and $S_\pi(t)$, whose suppression scales monotonically with $\alpha$ (panel~f).
}
\label{fig:fig3}
\end{figure*}

We systematically explore a range of phonon frequencies $\omega$ and pump fluences $\alpha$ while keeping $g_q/\omega$ fixed. The role of phonon frequency is critical: Small $\omega$ corresponds to slow phonon modes that excite large phonon populations, increasing the entanglement entropy and the amplitude of the disorder potential, yet simultaneously suppressing the width of the phonon number distribution—thereby reducing dynamical dephasing.

Our analysis of large system sizes allows us to resolve the momentum-dependent development of long-range correlations over time and under varying driving parameters—crucial for uncovering the nonlocal structure of spatial coherence, as we demonstrate in the following sections.

\begin{figure*}%[hbt]
\centering
\includegraphics[width=1.9\columnwidth]{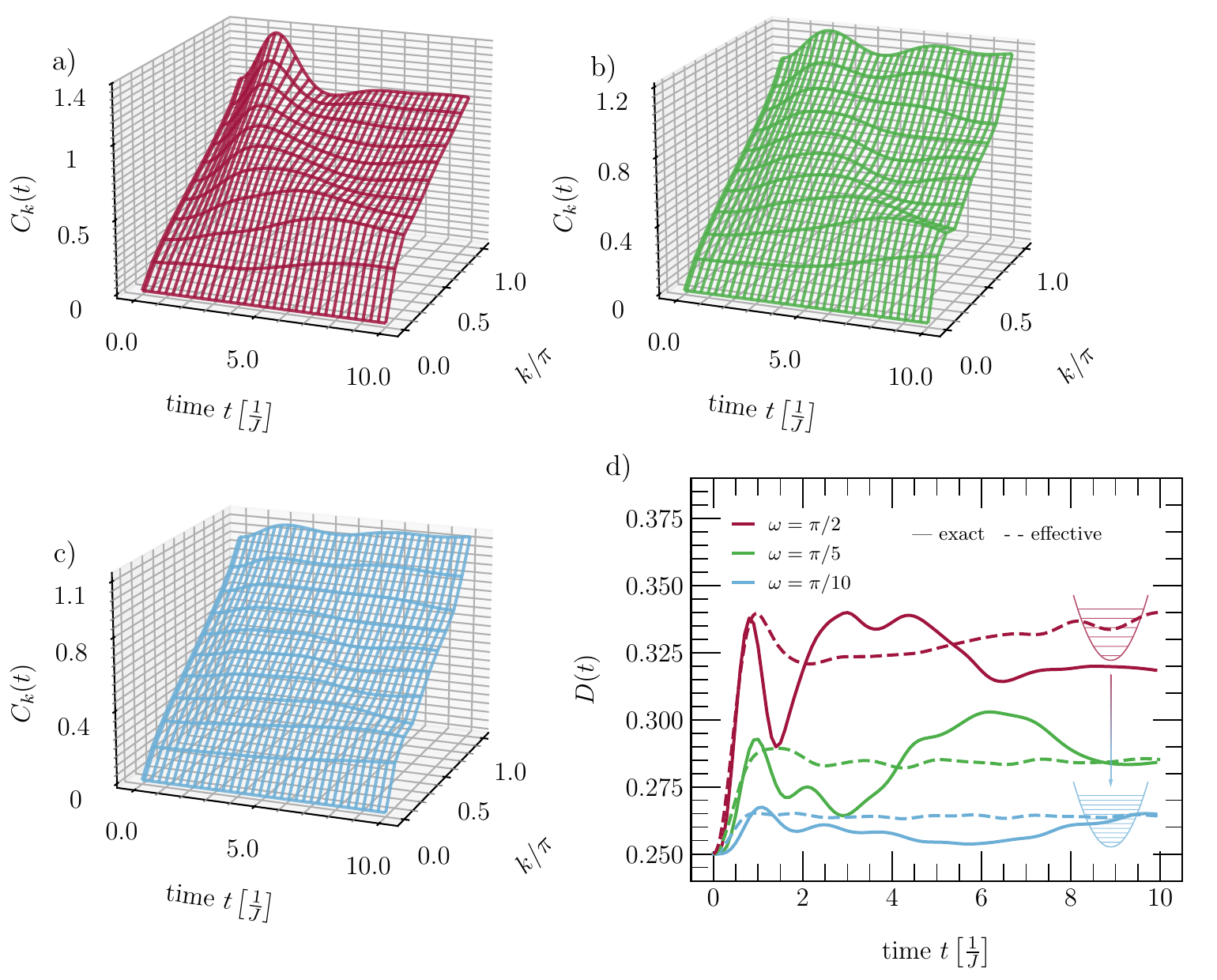}
\caption{\textbf{Effective model of light-enhanced long-range coherence via diminished disorder.} We analyze the dynamics of the momentum-resolved charge correlations $C_k(t)$ using the effective Hamiltonian in \cref{eq:effective:hamiltonian}, shown for decreasing phonon frequency from panel~a (largest) to panel~c (smallest), with parameters $\omega = (\pi/2)J$, $(\pi/5)J$, and $(\pi/10)J$ in panels~a, b, and c, respectively, and corresponding couplings $g_q = 0.35J$, $0.14J$, and $0.07J$, such that the ratio $g_q/\omega$ remains fixed. Panel~d compares the electronic double occupancy $D(t) \equiv \bra{\Psi(t)} \frac{1}{L} \sum_i \hat{n}_{i,\uparrow} \hat{n}_{i,\downarrow} \ket{\Psi(t)}$ obtained from the effective model with results from the full microscopic Hamiltonian~\cref{Eq:QephHam}. Time-integrated double occupancy shows similar agreement (not shown). All simulations were performed on systems of size $L=20$ sites. Notice that the effective model qualitatively captures key features: The flattening of $C_k(t)$ at high phonon frequency, indicative of dominant dynamic disorder, and the enhanced coherence at $k = \pi$ for low phonon frequency—although the latter is somewhat underestimated. These results support the utility of the effective model as a framework for understanding the competition between phonon-induced disorder and the emergence of long-range coherence. %All simulations were performed with $g_q = 0.07J$ on systems of size $L=20$ sites.
}
\label{fig:fig4}
\end{figure*}

\bigskip

\noindent\textbf{Light-mediated long-range pairing coherence}

\noindent  We begin by exploring the experimentally relevant and theoretically challenging regime of slow phonons, reflecting the large separation of energy scales between lattice vibrations and electronic dynamics in real materials. We study the regime of $\omega/J = \pi/10 \approx 0.3$ typical of most materials (e.g. transition-metal oxides) in Figure~\ref{fig:fig2}. The figure shows the time evolution of momentum-resolved charge ($C_k(t)$), pairing ($P_k(t)$), and spin ($S_k(t)$) correlations, characterizing the non-equilibrium electronic dynamics following an impulsive delta-function in time excitation that promptly and uniformly excites the lattice.

Charge correlations at momentum $\pi$ exhibit rapid initial enhancement with pronounced oscillations, followed by a transient dip and a significant upturn at later times ($t=12/J$), amounting to a $\sim20\%$ net increase. Pairing correlations show a brief initial suppression but subsequently recover and begin increasing in tandem with the growing charge correlations. By contrast, spin correlations are strongly suppressed overall, with only minor fluctuations during intermediate times. This combination—a growing pairing and charge response accompanied by suppressed spin coherence—suggests the emergence of long-range electronic correlations and pairing tendencies in the driven state. Importantly, since charge-density wave order tends to dominate over superconductivity in 1D half-filled systems~\cite{Huscroft1997, Assmann2012}, and the underlying mechanism due to slow phonons is not inherently tied to dimensionality, our results suggest that the enhancement of long-range coherence is poised to give rise to superconductivity in higher dimensions.

To understand the microscopic origin of this enhancement, we examine the momentum-resolved correlations across different phonon frequencies and pump fluences (Figure~\ref{fig:fig3}). For a  pump fluence corresponding to the excitation of $\langle n_{\mathrm{ph}} \rangle = \abs{\alpha}^2 = 2$ phonons per site, we observe three distinct dynamical regimes: (i) At high phonon frequencies, charge correlations exhibit a brief initial rise followed by rapid decay and subsequent flattening over momentum, indicative of disorder-dominated dynamics; (ii) At intermediate frequencies, clear oscillatory structure develops, signaling partial coherence; (iii) For low frequencies, correlations increase strongly over time, with a distinct enhancement at $k=\pi$, consistent with a developing staggered density order. These observations highlight a crucial physical mechanism: A strong separation of timescales between electronic and lattice degrees of freedom favors the emergence of long-range coherence.

In the lower panels of Figure~\ref{fig:fig3}, we fix $\omega = \pi/10$, the smallest value we study and vary the pump fluence $\alpha$. We observe that stronger values of the fluence monotonically enhance the $\pi$-momentum charge and  $0$-momentum (uniform) pairing correlations while suppressing the $\pi$-momentum spin order. This systematic dependence confirms that the growth of long-range electronic order is tunable by both energy scale and excitation strength.

\bigskip

\noindent\textbf{Effective model: Weakening disorder and enhanced coherence via slow phonons.}

\noindent To capture the physical origin of these effects, we derive an effective Hamiltonian valid in the weak-coupling, low-frequency regime via a canonical transformation into a squeezed phonon basis~\cite{Kennes2017,Sous2021} followed by a weak-coupling expansion. To second order in $g_q/\omega$, the electron and phonon dynamics decouple, yielding:
\begin{equation}
\begin{aligned}
\hat{H}_\text{eff} = & -J_\text{eff} \sum_{j,\sigma} \left( \hat{c}^\dagger_{j,\sigma} \hat{c}_{j+1,\sigma} + \text{h.c.} \right) \\
&+ U_\text{eff} \sum_j \left( \hat{n}_{j,\uparrow} - \tfrac{1}{2} \right)\left( \hat{n}_{j,\downarrow} - \tfrac{1}{2} \right) \left( \hat{b}_j^\dagger \hat{b}_j + \tfrac{1}{2} \right) \\
&+ \epsilon_\text{eff} \sum_j (\hat{n}_j - 1)\left( \hat{b}_j^\dagger \hat{b}_j + \tfrac{1}{2} \right)
+ \omega_\text{eff} \sum_j \left( \hat{b}_j^\dagger \hat{b}_j + \tfrac{1}{2} \right),
\end{aligned}
\label{eq:effective:hamiltonian}
\end{equation}
where $J_{\rm eff} = J e^{-\frac{1}{2}(g_q/\omega)^2(\alpha^4 + 2\alpha^2 + 1)}$, $U_{\rm eff} = -2g_q^2/\omega$, $\epsilon_{\rm eff} = 2g_q$, and $\omega_{\rm eff} = \omega - g_q^2/\omega$.

\begin{figure*}%[hbt]
\centering
\includegraphics[width=1.5\columnwidth]{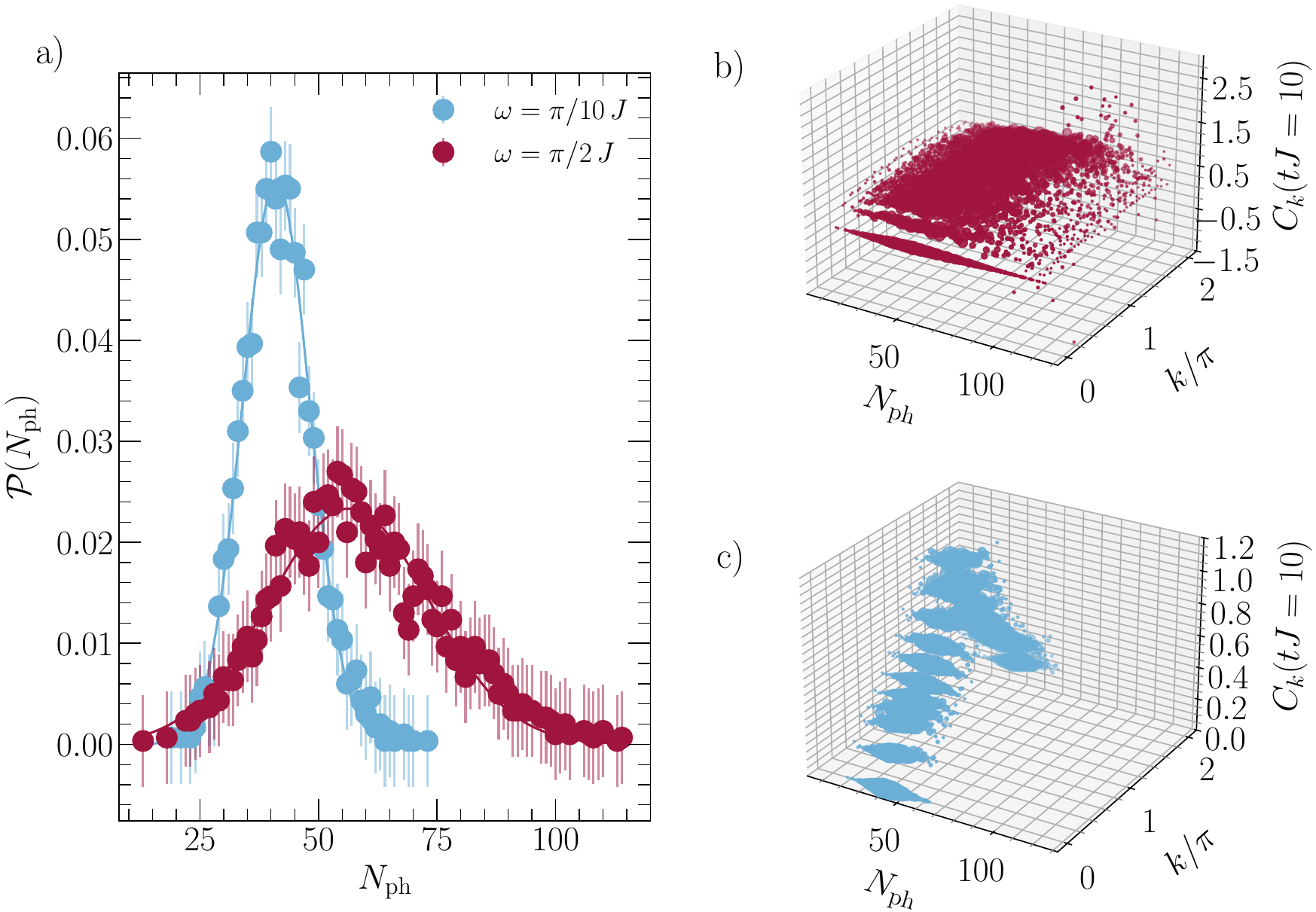}
\caption{\textbf{Long-time global phonon distribution for slow versus fast phonons.} Panel a shows the distribution function $\mathcal{P}(N_{\mathrm{ph}})$ of total phonon quanta excited across the entire system $N_{\mathrm{ph}}$ at late times ($tJ = 10$). We find that $\mathcal{P}(N_{\mathrm{ph}})$ is substantially narrower for low phonon frequency (blue dots) than for high phonon frequency (red dots). This narrowing implies that the decomposition of the charge correlation $C_k(tJ = 10)$ into contributions from different total phonon number sectors $N_{\mathrm{ph}}$ spans a wider range for high-frequency phonons (panel b) than for low-frequency ones (panel c). Results correspond to a system of size $L = 20$, obtained using $3000$ PST samples. These results highlight how the phonon energy scale governs the width of the disorder distribution.
}
\label{fig:fig5}
\end{figure*}

Figure~\ref{fig:fig4} compares the charge correlation dynamics from the full quantum simulation and the effective model. The effective model qualitatively captures key features—such as the flattening at high $\omega$, indicative of enhanced disorder, and the peak at $k = \pi$ for low $\omega$, signaling the growth of long-range coherence—and also semi-quantitatively reproduces local charge observables like the double occupancy $D(t)$.  Coupled with the effective model, we analyze the full distribution of the total phonon excitations, $P(N_{\rm ph})$, obtained from the global phonon reduced density matrix using phonon state tomography (PST)~\cite{moroder2024}. As illustrated in Figure~\ref{fig:fig5}, at a given total phonon number excited globally over the entire system, the distribution $P(N_{\rm ph})$ is substantially narrower at low $\omega$ than at high $\omega$, indicating reduced phonon-induced disorder. Decomposition of $C_k(t)$ by phonon number sector shows more structured, coherent contributions at low $\omega$ and flat, unstructured behavior at high $\omega$. See Supplementary Materials for more details.

Taken together, a key physical picture emerges: Slow phonons produce narrow phonon distributions, minimizing decoherence and enabling the emergence of long-range order. We can understand this through the effective model. In the transformed squeezed basis, the nonlinear electron-phonon coupling gives rise to a disorder potential of strength $\epsilon_{\mathrm{eff}}$. This disorder, while not quenched, leads to decoherence through a mechanism akin to disorder-free localization~\cite{Smith2017,Smith2017-2,Halimeh2022,Buessen2022}. Here, even in the absence of static randomness, dynamical processes generate effective disorder. As highlighted in Ref.~\cite{Sous2021}, this occurs because in the squeezed frame the phonon occupations become conserved quantities—constants of motion—and observables evolving over time are a sum over contributions from fixed phonon-number sectors. This structure maps formally onto the Anderson localization problem (in absence of electron-electron interactions), where the dynamics is averaged over static disorder realizations, except that in our case the averaging arises dynamically from the different phonon number contributions. \emph{However}, the above analysis misses a crucial subtlety: As the phonon frequency $\omega$ decreases, the energy spacing between phonon number states in the Einstein oscillator shrinks (illustrated in panel d of Figure~\ref{fig:fig4}). In the limit $\omega \to 0$, corresponding to classical phonons, the energy differences vanish and the Poisson-distributed initial phonon population created by the pump effectively collapses onto a single number state. Thus, the self-averaging over phonon sectors disappears, and the effective disorder vanishes. For small but finite $\omega$, a similar tendency emerges—the phonon distribution narrows, reducing disorder and enabling long-range coherence to build up. Nevertheless, a finite phonon energy implies that some residual disorder remains, which may induce disorder-free localization at long times—even after long-range correlations have developed. This points to a potentially exotic scenario: The phonon energy scale may set a dynamical hierarchy in which coherence emerges early, while localization stabilizes the resulting non-equilibrium phase against thermalization at later times, ultimately giving way to a fluctuating state.

Together, these findings confirm that light-induced long-range charge correlations are mediated not simply by the pump amplitude, but by the interplay between the phonon energy scale, excitation fluence, and electron-phonon coupling. Our results suggest that narrowing the phonon distribution may be a necessary condition for realizing transient coherence, offering a concrete criterion for experimental design of light-induced phases.

\bigskip

\noindent\textbf{Conclusion}

\noindent In this work, we investigated the dynamics of a 1D  metal with nonlinear electron-phonon coupling subjected to an impulsive optical pump. Using a numerically exact MPS approach, we accessed the realistic regime of slow phonons and strong excitation, enabling simulations of long-time dynamics in systems with large local Hilbert spaces and large system sizes. Our model captures the essential physics of light-driven dipole-active phonon modes and their coupling to electronic degrees of freedom. We demonstrated that even modest enhancements in local pairing are sufficient—when coupled to slow phonons—to generate significant long-range coherence in charge and pairing correlations. This provides the first controlled, non-perturbative evidence of light-induced enhancement of long-range phase ordering tendencies mediated by phonons.

Our findings offer new insights into the mechanisms behind light-enhanced superconductivity and non-equilibrium ordering. We identified a key role for the phonon energy scale: Slow phonons yield narrower phonon number distributions resulting in a reduction of the effective disorder, enabling coherence to grow dynamically. This challenges conventional assumptions that stronger pump fluence alone is the route to enhanced ordering, and instead highlights the importance of timescale separation between electronic and lattice degrees of freedom. These results establish a framework for designing ultrafast protocols that leverage low-energy vibrational modes to stabilize correlated electronic phases far from equilibrium.

In experiment, the mechanisms by which optical phonon excitations enhance superconducting correlations remain under active debate. One proposed scenario is that high-frequency modes couple nonlinearly to low-energy, symmetry-breaking lattice distortions. For instance, in YBa$_2$Cu$_3$O$_x$, nonlinear phononics has been shown to transiently modulate the apical oxygen position~\cite{Forst2011,Mankowsky2014nonlinear,Mankowsky2015coherent}, while in K$_3$C$_{60}$, high-frequency intramolecular modes may couple to low-energy Jahn-Teller radial distortions thought to contribute to pairing. Recent observations of a resonant enhancement of superconductivity near 40~meV in K$_3$C$_{60}$~\cite{Rowe2023resonant} further support the involvement of low-energy phonons in light-induced phenomena. This, combined with our findings, motivates new experimental directions. Our results suggest that low-energy phonons can play an active role in driving long-range coherence. Quasi-1D systems are particularly well suited to test this hypothesis, yet have received comparatively less attention than two- and three-dimensional systems. Theoretical proposals suggest that driven 1D correlated systems may host exotic non-equilibrium orders such as $\eta$-pairing superconductivity~\cite{Kaneko2019,Li2020,Peronaci2020,Murakami2023}, with candidate materials including doped chain cuprates Ba$_{2-x}$Sr$_x$CuO$_{3+\delta}$~\cite{Chen2021anomalously,Li2025doping} and ladder compounds like Sr$_{14-x}$Ca$_x$Cu$_{24}$O$_{41}$~\cite{Padma2025beyond, Padma2025}, where low-energy Raman-active or sliding phonon modes~\cite{Thorsmolle2012phonon} are prominent. Another promising platform is provided by quasi-1D metals such as (TMTSF)$_2$PF$_6$, where intramolecular vibrational modes of TMTSF can couple to lower-frequency lattice phonons. These systems offer a clean setting to test whether coherent control over low-energy phonon populations can trigger long-range superconducting correlations, in line with the theoretical mechanism demonstrated here.

Furthermore, the PST-simulated global phonon distributions provide a pathway for probing phonon coherence in pump-probe experiments~\cite{moroder2024}, which may be complemented by techniques for detecting electronic coherence~\cite{Paeckel2020,grundner2023cooperpaired}. Time-resolved x-ray or electron diffraction can measure the time-dependent diffuse scattering intensity which—together with knowledge of the phonon frequency and the unperturbed structure factor—allows extraction of the phonon occupation distribution~\cite{Tollerud2019,Glerean:20,glerean2025}. This distribution can then be directly compared to simulation outputs from PST, enabling a physically grounded interpretation of the experimental data.

Looking forward, our results suggest several promising directions. One key insight is that disorder suppression plays a crucial role in enabling long-range coherence. In our model, this was achieved by utilizing slow phonons, which naturally narrow the distribution of phonon-induced potentials. However, this outcome may also be engineered through tailored pump protocols. Specifically, one could design pulses that selectively populate a single or few excited phonon states—rather than a thermal-like distribution—to mimic the effective narrowing achieved by slow phonons. In particular, pulse schemes inspired by $\pi$-pulses~\cite{Cao1998, Dlott1990}, which can selectively transfer phonon population into a specific excited state, offer a route to minimize pump-induced disorder without requiring extremely low phonon energies. By sharply narrowing the phonon distribution, such protocols could enhance phase coherence and enable the stabilization of high-temperature, transient ordered states in driven quantum materials.

Ultimately, our study provides both a rigorous demonstration of light-enhanced long-range order in a minimal microscopic model, and a conceptual framework to guide future experimental and theoretical efforts in non-equilibrium phase control.

\FloatBarrier
\bibliography{Literature}

\bigskip
\noindent{\textbf{Methods}}

\noindent We evolve the initial metallic state in time under the full Hamiltonian, Eq.~\eqref{Eq:QephHam}, following an impulsive excitation of its phonons, using large-scale MPS simulations. Separately, we evolve the same initial state under the effective model, Eq.~\eqref{eq:effective:hamiltonian}. The effective model can be derived within a framework analogous to linear response theory~\cite{Sous2021}, systematically incorporating contributions of order $\mathcal{O}(g_q/\omega)$, along with carefully selected terms of order $\mathcal{O}((g_q/\omega)^2)$. To overcome limitations of previous studies—such as two-site models~\cite{Kennes2017,Sentef2017,eckhardt2024} or conventional MPS methods with restricted local Hilbert spaces~\cite{Sous2021}—we employ the projected purification (PP) framework for bosonic states~\cite{Koehler2021}, together with the linear subspace expansion time-dependent variational principle (LSE-TDVP)~\cite{GSE,grundner2023cooperpaired}.  This enables us to capture exact dynamics in systems with up to $L=32$ sites and phonon cutoff dimension $d_\nu=40$, accessing regimes with slow phonons and long evolution times ($t \sim 12/J$) that were previously inaccessible. The pump-induced dynamics are computed with high numerical precision. We use a maximal bond dimension $\chi_{\mathrm{max}} = 2000$ and enforce a local truncation error $\epsilon_{\mathrm{ph}} = \epsilon_{\mathrm{1s\mbox{-}MPS}} = 10^{-10}$ throughout.  The results in this work were produced using SyTen~\cite{hubig:_syten_toolk,hubig_thesis}. Details of convergence checks are reported in the Supplementary Materials.

\bigskip
\noindent{\textbf{Acknowledgments}}

\noindent M. Moroder and S. Paeckel acknowledge support by the Deutsche Forschungsgemeinschaft (DFG, German Research Foundation) under Germany’s Excellence Strategy-426 EXC- 2111-390814868.  M. Moroder also acknowledges funding from the Royal
Society and Research Ireland. M. Mitrano acknowledges support by the U.S. Department of Energy, Office of Basic Energy Sciences, Early Career Award Program, under Award No. DE-SC0022883. J. Sous acknowledges support from the Air Force Office of Scientific Research under Award No. 14191224. The authors wish to thank A. Cavalleri, T. Grover, A.~J. Millis, P.~A. Volkov, and J. Yuen-Zhou for insightful discussions.

\end{document}

\author{M.~Moroder}
\affiliation{School of Physics, Trinity College Dublin, College Green, Dublin 2, D02K8N4, Ireland}
\author{S.~Paeckel}
\affiliation{Department of Physics, Arnold Sommerfeld Center for Theoretical Physics (ASC), Munich Center for Quantum Science and Technology (MCQST), Ludwig-Maximilians-Universit\"{a}t M\"{u}nchen, 80333 M\"{u}nchen, Germany}
\author{M.~Mitrano}
\affiliation{Department of Physics, Harvard University, Cambridge, Massachusetts 02138, USA}
\author{J.~Sous}
\affiliation{Department of Applied Physics, Yale University, New Haven, Connecticut 06511, USA}
\affiliation{Energy Sciences Institute, Yale University, West Haven, Connecticut 06516, USA}

\def\thetitle{Supplementary Information for: \\``Ultrafast electronic coherence from slow phonons"} 
\title{\thetitle}
\maketitle

\section{\label{app:sec:convergence} Methods}
\begin{figure}
    \centering
    \includegraphics[width=0.85\linewidth]{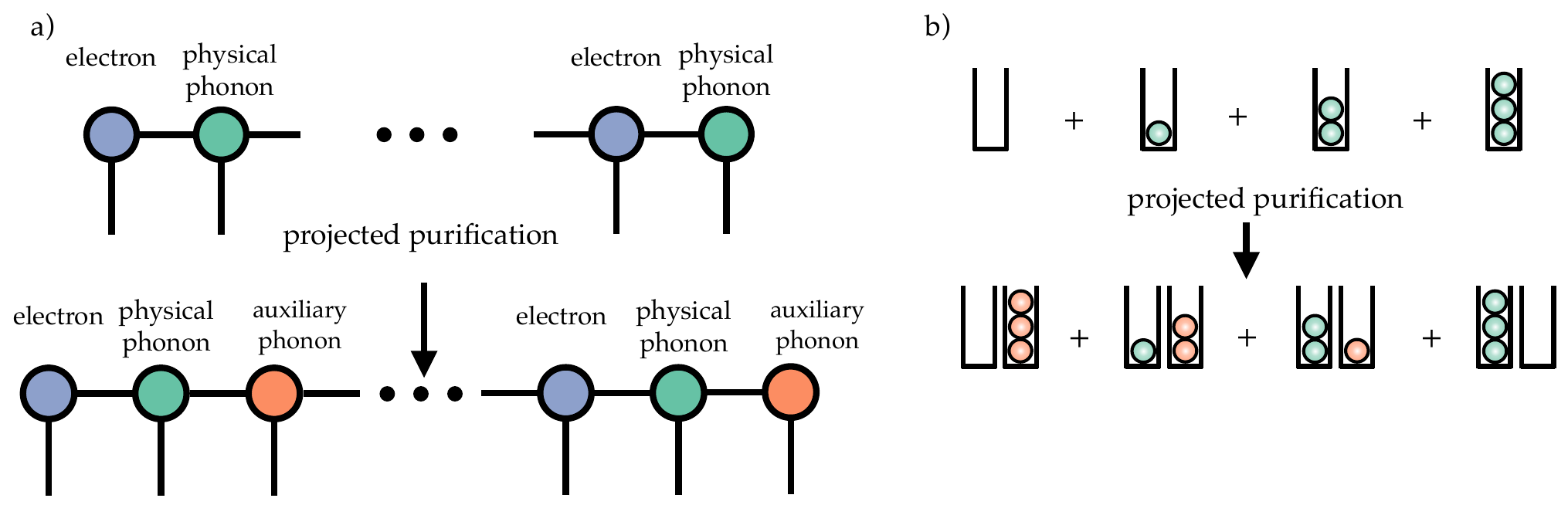}
    \caption{Panel a): The electron-phonon state (see \cref{eq:psi:for:pst}) is represented as an \gls{MPS} of alternating electronic and phononic sites. We adopt the \acrfull{PP} representation \cite{Koehler2021}, which implements  doubling of the number of phononic sites. Panel b): Illustration of the projected purification (PP) mapping. For clarity, we show a single truncated bosonic mode with four levels (top). In the \gls{PP} mapping (bottom), auxiliary sites are introduced to enforce conservation of the total boson number.
    }
    \label{fig:app:schematic:pp}
\end{figure}

Below, we detail our simulation methods and verify their numerical convergence across all parameter regimes examined.

\subsection{Formalism}

We encode the entire electron-phonon wavefunction as a matrix product state (\gls{MPS})~\cite{schollwoeck_2011}, in which electronic and phononic sites alternate, as illustrated in Figure~\ref{fig:app:schematic:pp}a). We initialize the system in its ground state, which we verify numerically to be very weakly entangled and equivalent to a product state of electrons in a Fermi sea and a coherent state of phonons
\begin{equation}
    \ket{\Psi_0} = \ket{\mathrm{FS}}\otimes \ket{\alpha} \,,
    \label{eq:supp:mat:initial:state}
\end{equation}
where $\ket{\mathrm{FS}} = \prod_{k \in \left([0, \frac{\pi}{2}] \cup [\frac{3\pi}{2}, 2\pi]\right)} \hat{c}_{k,\uparrow}^\dagger \hat{c}_{k,\downarrow}^\dagger \lvert 0 \rangle^\text{el}$ and $\ket{\alpha} = \otimes_{i=1}^L \ket{\alpha_i}$.
Here, $\hat{c}_{k,\uparrow}^\dagger$ ($\hat{c}_{k,\downarrow}^\dagger$) creates a spin up (spin down) fermion with momentum $k$, the index $i$ labels the lattice sites, and $L$ is the total number of sites.
The coherent state on each site $\ket{\alpha_i}$ is obtained by applying the displacement operator $\hat{\mathcal{D}}(\alpha) =  \prod_i e^{\alpha \hat{b}^\dagger_i - \alpha^* \hat{b}_i}$ on the vacuum state, where $\hat{b}^\dagger$ ($\hat{b}$) creates (annihilates) a bosonic excitation. 
The Hamiltonian is given by
\begin{equation}
    \hat{H} = \underbrace{-J \sum_{i,\sigma} \hat{c}^\dagger_{i,\sigma} \hat{c}_{i+1,\sigma} + \mathrm{H.c.}}_{\hat{H}_e} + \underbrace{\omega \sum_i \left( \hat{b}^\dagger_i \hat{b}_i + \tfrac{1}{2} \right)}_{\hat{H}_\mathrm{ph}} + \underbrace{g_q \sum_i (\hat{n}_i - 1)(\hat{b}^\dagger_i + \hat{b}_i)^2}_{\hat{H}_\mathrm{e-ph}}
    \label{app:eq:total:hamiltonian}
\end{equation}
where $\hat{H}_e$ describes spinful tight-binding electrons with hopping $J$, $\hat{H}_\mathrm{ph}$ Einstein oscillators with frequency $\omega$ and $\hat{H}_\mathrm{e-ph}$ a non-linear local coupling between the electron density $\hat{n}_i = \sum_\sigma \hat{c}^\dagger_{i,\sigma} \hat{c}_{i,\sigma}$ and the oscillator's displacement with interaction strength $g_q$.
The time evolution generated by \cref{app:eq:total:hamiltonian} (which is represented as a \gls{MPO} \cite{HubigMPO2017}) is computed using the~\gls{TDVP}~\cite{Haegeman2011} for~\gls{MPS}'s~\cite{time_ev_methods_mps}.
To mitigate the projection error~\cite{time_ev_methods_mps}, which is particularly severe for product states such as \cref{eq:supp:mat:initial:state}, we compute the first four steps of the time evolution with the~\gls{GSE-TDVP}~\cite{GSE}.
Then we switch to the faster~\gls{LSE-TDVP}~\cite{GSE,grundner2023cooperpaired}.
The two most important parameters which determine the numerical cost of the simulations are the bond dimension $\chi$ \cite{schollwoeck_2011}, and the local Hilbert space dimension of the phononic sites $d_\nu$. 

A key ingredient enabling these simulations is the projected purification (PP) representation~\cite{Koehler2021}, which is crucial for reaching large system sizes and long simulation times. PP enables us to utilize an artificial $\mathrm{U}(1)$ symmetry corresponding to the conservation of the total number of real and artificial bosons, even though the (nonlinear) electron-phonon coupling $\propto (\hat c^\dagger_i\hat c^\nodagger_i-1)(\hat b^\dagger_i + \hat b^\nodagger_i)^2$ does not conserve the number of real boson excitations. This is implemented by first doubling the number of bosonic sites, introducing auxiliary partners for each physical phonon site, and then projecting the enlarged Hilbert space onto the subspace where the total number of bosons on each physical--auxiliary pair is fixed to $d_\nu^{\mathrm{max}} - 1$. This projected Hilbert space is isomorphic to the original (non-purified) Hilbert space and therefore introduces no additional interactions, as represented pictorially in \cref{fig:app:schematic:pp}b). Importantly, the artificial $\mathrm{U}(1)$ symmetry enables a decomposition of the site tensors into smaller symmetry blocks, accelerating the time evolution by roughly a factor of~$d_\nu$. 
Furthermore, the physical local dimension $d_\nu$ is dynamically adjusted during the simulation~\cite{Koehler2021}, just as the bond dimension $\chi$ is in standard \gls{DMRG} algorithms~\cite{schollwoeck_2011}, providing an additional speed-up.

\subsection{Numerical accuracy}
During the time evolution within the \gls{PP} framework~\cite{Koehler2021}, the growth of both the bond dimension $\chi$ and the local phonon Hilbert space dimension $d_\nu$ is controlled by the discarded weight $\epsilon$, which measures the information lost in each truncated \gls{SVD}\,\cite{time_ev_methods_mps}.
In all simulations reported in the main text, we cap the local phonon dimension at $d_\nu^{\max}=40$ and the bond dimension at $\chi_{\max}=2000$. A uniform discarded weight of $\epsilon_{\text{1s--MPS}}=\epsilon_{\text{ph}}=\epsilon=10^{-10}$ is applied to both bond and physical truncations, and the bond dimension is permitted to grow by no more than 100 per time step of the \gls{LSE-TDVP} algorithm~\cite{grundner2023cooperpaired}.
The calculations were performed with the tensor networks toolkit \textsc{SyTen}~\cite{hubig:_syten_toolk, hubig17_symmet_protec_tensor_network}.
\begin{figure}[!b]
    \centering
    \includegraphics[width=0.95\linewidth]{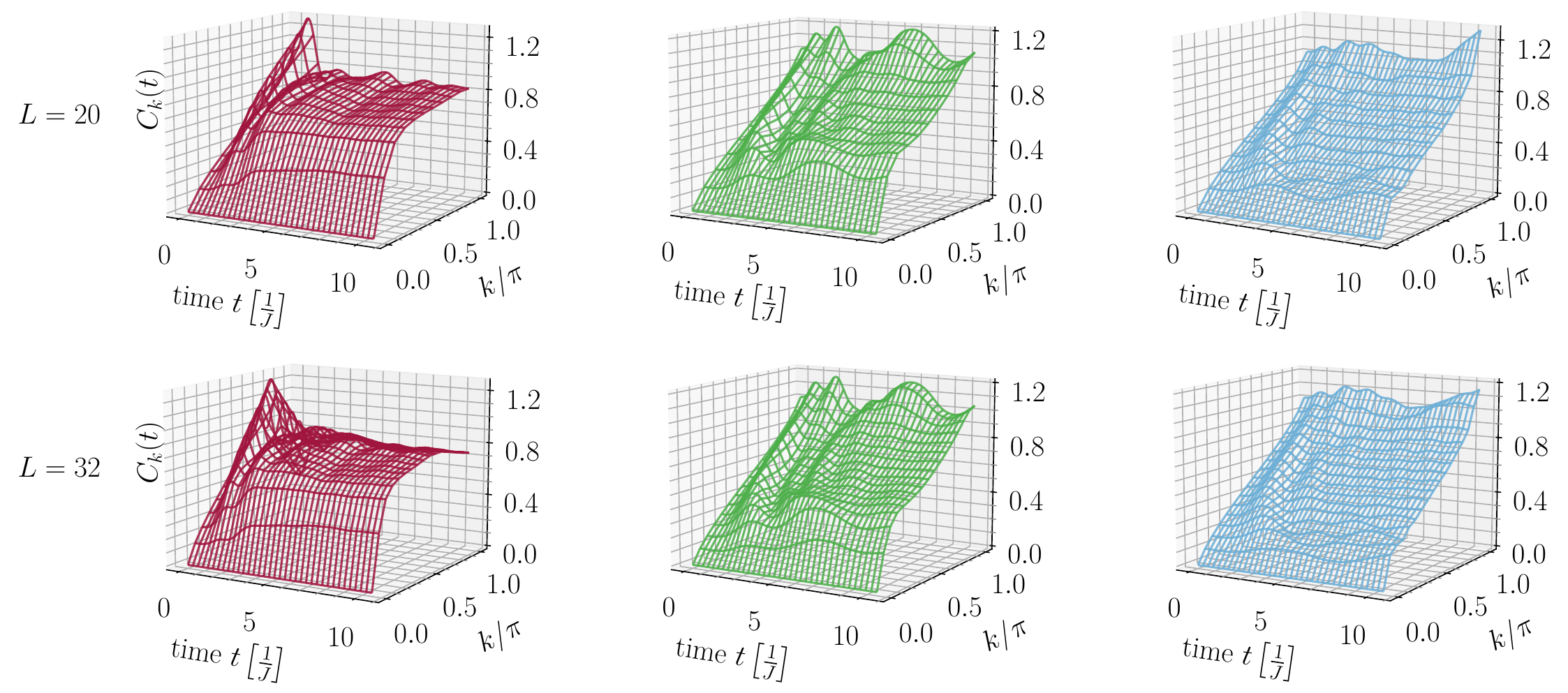}
    \caption{Dynamics of the momentum\hyp resolved charge correlations for $L=20$ sites (upper row) and $L=32$ sites (lower row). Here, we show the Fourier transform $\mathcal F$ of the charge correlation function, $C_k(t) \equiv \mathcal{F}[\bra{\Psi(t)} \hat{n}_i \hat{n}_{i+r} \ket{\Psi(t)} - \bra{\Psi(t)} \hat{n}_i\ket{\Psi(t)} \bra{\Psi(t)} \hat{n}_{i+r}\ket{\Psi(t)}]$ for a large value of the phonon frequency ($\omega=(\pi/2) J$, left panels), an intermediate value of the phonon frequency ($\omega=(\pi/5) J$, central panels) and a small value of the phonon frequency ($\omega=(\pi/10) J$, right panels), with $g_q / \omega = 0.7/\pi \approx 0.22$ and $\alpha = \sqrt{2}$. We set $d_\nu^{\mathrm{max}}=40$ and $\chi_\mathrm{max}=2000$ in these simulations.}
    \label{fig:app:Ck:L20:L32}
\end{figure}
In~\cref{fig:app:Ck:L20:L32} we show the dynamics of the momentum-resolved charge correlations for system sizes $L=20$ and $L=32$. We can see that semi-quantitative convergence is achieved, while the key features—the flattening of correlations at high phonon frequencies (left panels) and the late-time enhancement of $C_\pi$ (middle and right panels)— have essentially converged.  Next, we examine $k=\pi$ cuts computed with three bond-dimension ceilings, $\chi_{\max}=1500, 2000$, and $2500$ (see Figure~\ref{fig:app:bdim:conv}). The results confirm full convergence for both system sizes at low and intermediate phonon frequencies (blue and green lines) and show that late-time data at high phonon frequencies (red lines) differ by only a few percent, indicating near-convergence.
\begin{figure}[H]
    \centering
    \includegraphics[width=0.8\linewidth]{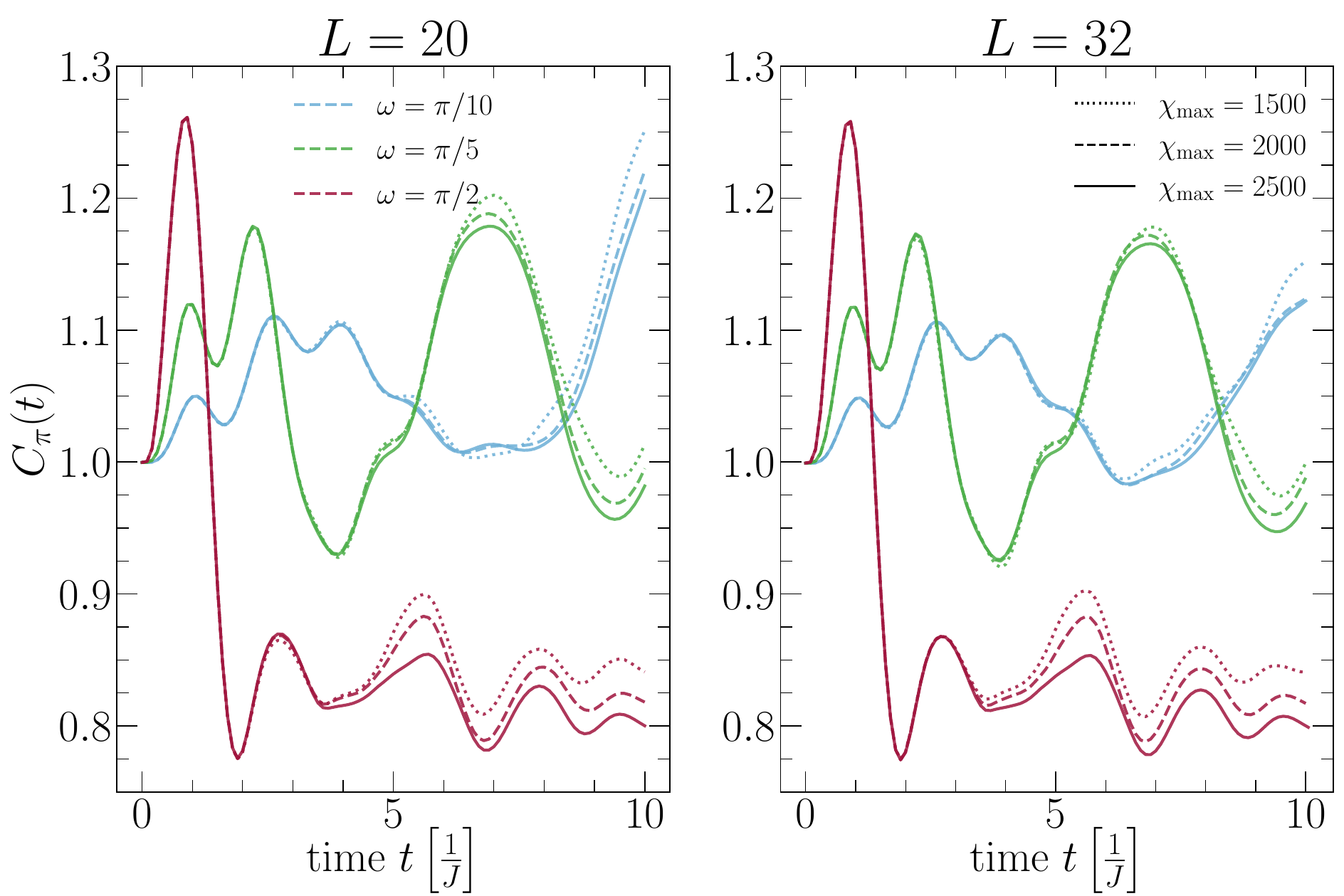}
    \caption{Convergence analysis of the charge correlation function at momentum $k=\pi$  with respect to the maximum allowed bond dimension  $\chi_\mathrm{max}$ for different phonon frequencies and system size $L=20$ (left panel) and $L=32$ (right panel). We use $g_q / \omega = 0.7/\pi \approx 0.22$ and $\alpha = \sqrt{2}$ in this figure.
    } 
    \label{fig:app:bdim:conv}
\end{figure}
To carefully gauge finite-size effects, we introduce the $k=\pi$ charge-density imbalance
\begin{equation}
\Delta(t)=\frac{1}{c}\sum_{i,r=s}^{L-s}(-1)^{|i-r|}C(i,r),\qquad 
C(i,r)=\langle\Psi(t)|\hat n_i\hat n_{i+r}|\Psi(t)\rangle-
\langle\Psi(t)|\hat n_i|\Psi(t)\rangle\,
\langle\Psi(t)|\hat n_{i+r}|\Psi(t)\rangle,
\label{eq:app:imbalance}
\end{equation}
where we average over all pairs that lie at least $s$ sites away from each boundary; the normalization is $c=(L-2s)\sqrt{L-2s}$.  
Figure~\ref{fig:app:imbalance} plots $\Delta(t)/\Delta(0)$ for the three phonon frequencies—large (red), intermediate (green), and small (blue)—with boundary exclusions $s = 2, 3, 4,$ and 5.  
This observable mirrors $C_\pi(t)$ (see the right panel of Figure~\ref{fig:app:bdim:conv}) and shows that finite-size effects are essentially eliminated at low phonon frequencies (blue curves).

\begin{figure}
    \centering
    \includegraphics[width=0.75\linewidth]{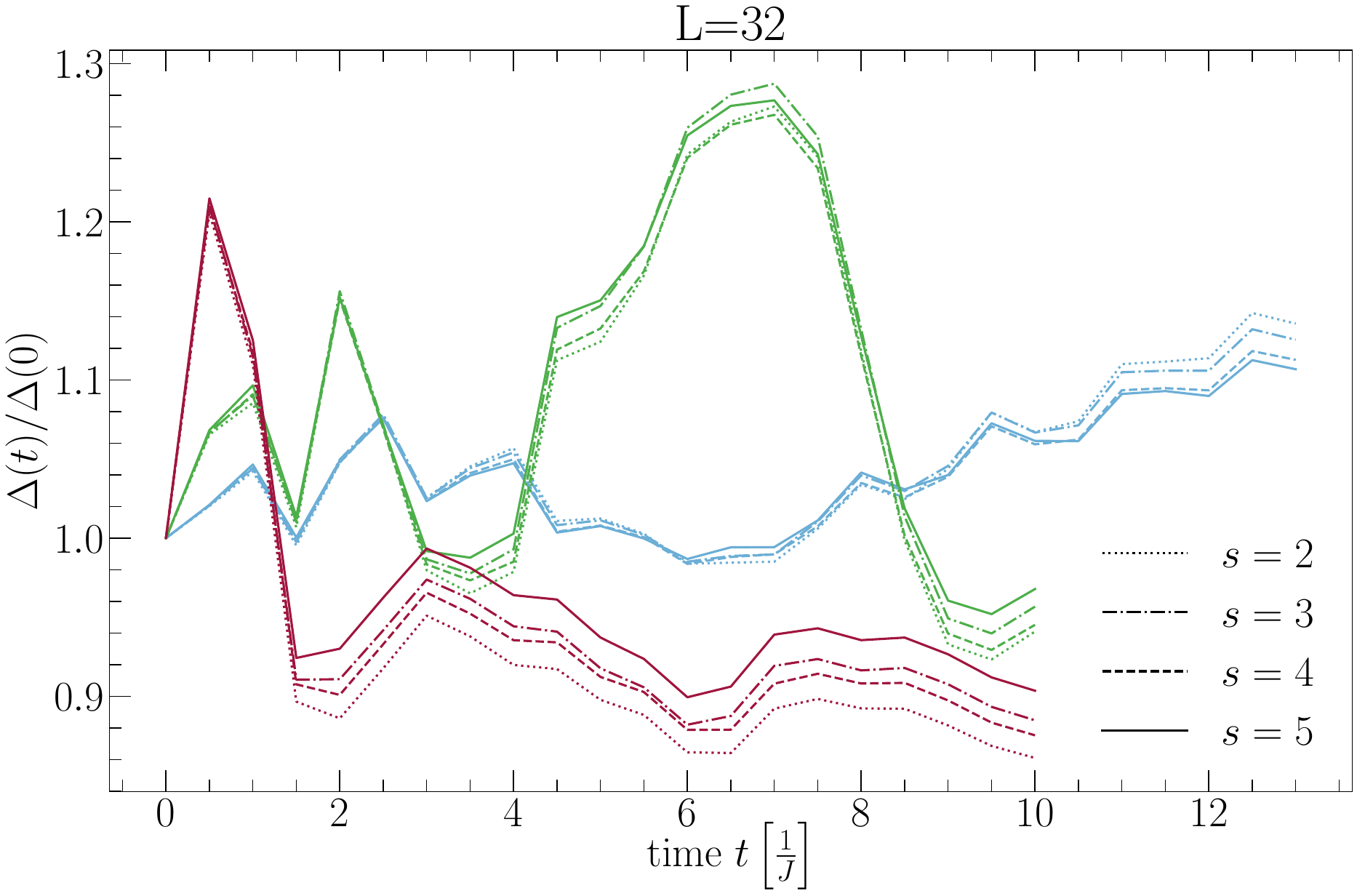}
    \caption{ Finite-size analysis of the $k=\pi$ charge imbalance $\Delta(t)/\Delta(0)$ for fast $(\omega=\pi/2J)$ (red), intermediate $(\omega=\pi/5J)$ (green), and slow $(\omega=\pi/10J)$ (blue) phonons.  Systematically removing $s$ boundary sites (cf.\ \cref{eq:app:imbalance}) shows that finite-size effects are nearly negligible at low phonon frequencies and remain milder than those at high frequencies.  All simulations use a chain length $L=32$, a maximum bond dimension $\chi_{\max}=2000$, and a phonon cutoff $d_\nu^{\max}=40$.  The coupling parameters are $g_q/\omega = 0.7/\pi \approx 0.22$ and $\alpha = \sqrt{2}$. 
    }
    \label{fig:app:imbalance}
\end{figure}

\section{\label{app:sec:effective:model} The effective model}
\begin{figure}
    \centering
    \includegraphics[width=\linewidth]{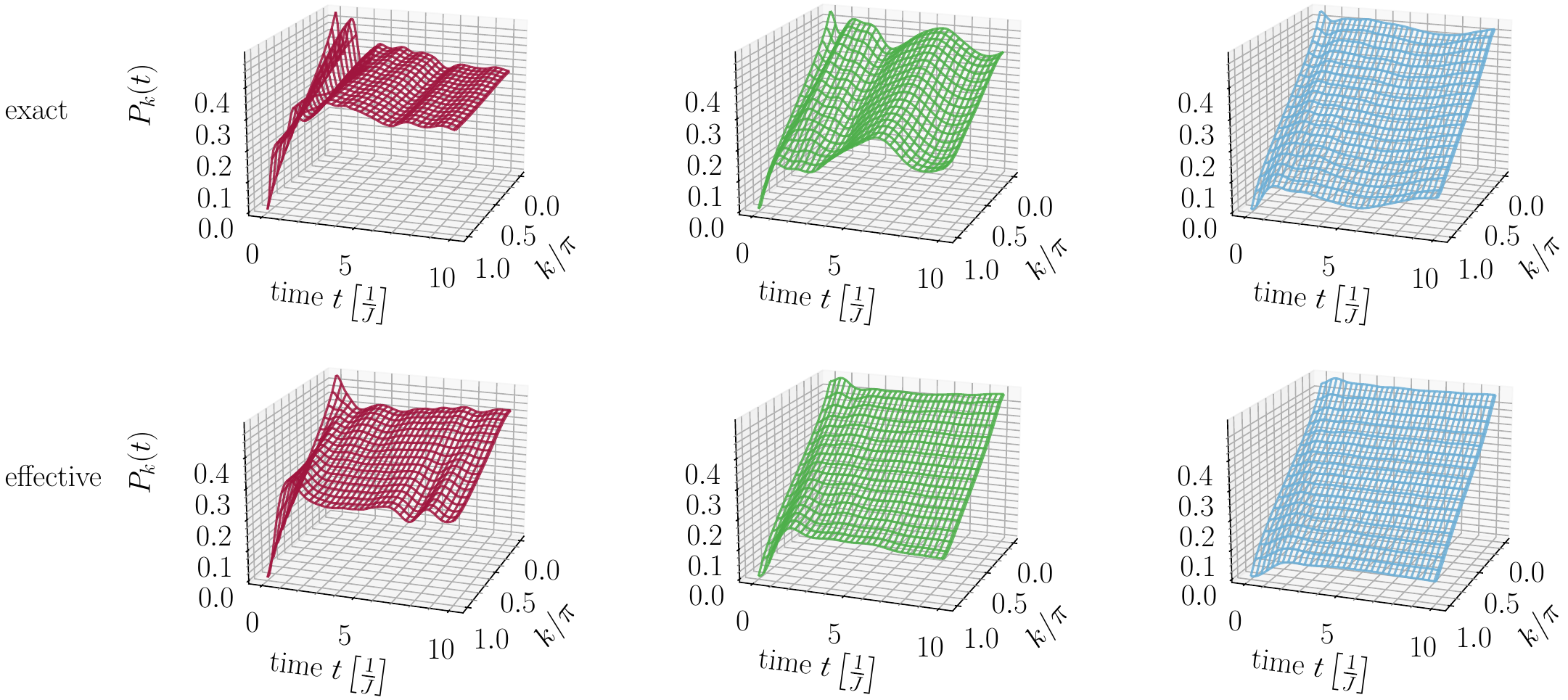}
    \caption{Comparison of momentum-resolved pairing correlations $P_k(t)\equiv\mathcal{F}\!\bigl[\langle\Psi(t)|\hat c_{i\uparrow}^\dagger \hat c_{i\downarrow}^\dagger \hat c_{i+r\downarrow}^{\phantom{\dagger}}\hat c_{i+r\uparrow}^{\phantom{\dagger}}|\Psi(t)\rangle\bigr]$ for the exact dynamics (top row) and the effective model (bottom row), where $\mathcal{F}$ denotes the Fourier transform.  Following Figure~\ref{fig:app:Ck:L20:L32}, we examine three phonon frequencies—fast ($\omega=\pi/2\,J$, left), intermediate ($\omega=\pi/5\,J$, centre), and slow ($\omega=\pi/10\,J$, right)—while keeping the ratio $g_q/\omega = 0.7\pi \approx 0.22$ fixed.  All simulations use a chain length $L=20$, a maximum bond dimension $\chi_{\max}=2000$, and a phonon cutoff $d_\nu^{\max}=40$.}
    \label{fig:supp:pairing:effective}
\end{figure}

\begin{figure}
    \centering
    \includegraphics[width=\linewidth]{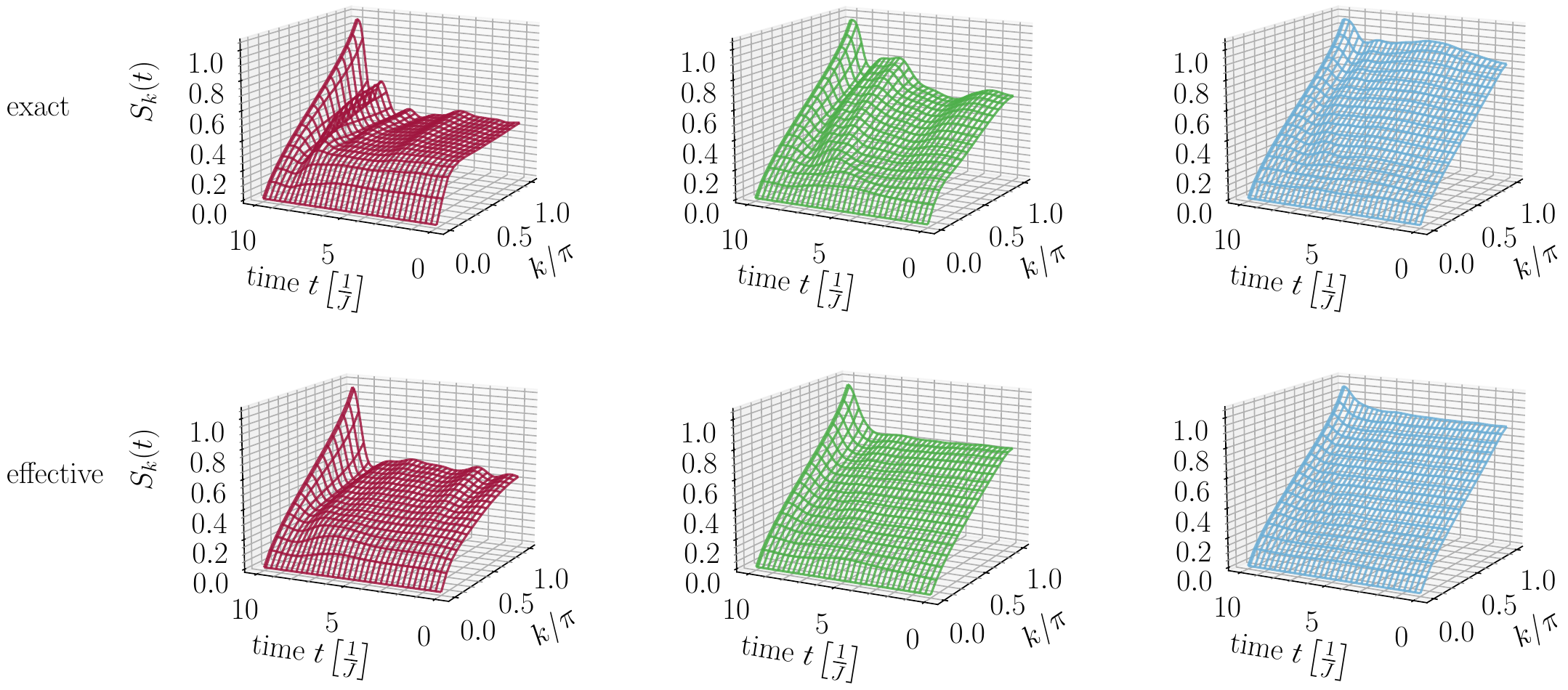}
    \caption{
    Comparison of momentum-resolved spin correlations $S_k(t)\equiv\mathcal{F}\!\bigl[\langle\Psi(t)|(\hat n_{i\uparrow}-\hat n_{i\downarrow})(\hat n_{i+r\uparrow}-\hat n_{i+r\downarrow})|\Psi(t)\rangle\bigr]$ for the exact dynamics (upper row) and the effective model (lower row), where $\mathcal{F}$ denotes the Fourier transform.  We examine three phonon frequencies—fast ($\omega=\pi/2\,J$, left), intermediate ($\omega=\pi/5\,J$, centre), and slow ($\omega=\pi/10\,J$, right)—while keeping the ratio $g_q/\omega = 0.7\pi \approx 0.22$ fixed.  All simulations use a chain length $L=20$, a maximum bond dimension $\chi_{\max}=2000$, and a phonon cutoff $d_\nu^{\max}=40$. }
    \label{fig:supp:spin:effective}
\end{figure}

In this section, we analyze in more detail the effective model studied in the main text. 
It was first introduced by Kennes et al.~\cite{Kennes2017} and further developed by Sous et al.~\cite{Sous2021}.
The main idea behind this approach is to rotate the exact Hamiltonian into the squeezed phonon basis, where the dynamics of electrons and phonons decouple. 
This rotation is obtained via the unitary operator $\hat{U} = e^{\hat{S}} = e^{-\sum_i \frac{1}{2}\zeta_i(\hat{b}_i^\dagger \hat{b}_i^\dagger - \hat{b}_i\hat{b}_i)}$, where the squeezing parameter is $\zeta_i = -\frac{1}{4} \ln(1+4\frac{g_q}{\omega}(n_i-1))$. 
The squeezed phonon creation and annihilation operators are related to original operators as $\hat \beta^\dagger_i \equiv e^{\hat S} \hat{b}^\dagger_i  e^{-\hat S} = \cosh(\zeta_i) \hat{b}^\dagger_i + \sinh(\zeta_i) \hat{b}_j$ and $\hat \beta_i \equiv e^{\hat S} \hat{b}_i  e^{-\hat S} = \cosh(\zeta_i) \hat{b}_i + \sinh(\zeta_i) \hat{b}^\dagger_i$.
The transformation $\hat{H} \to \hat{H}' = \hat{U}\hat{H}\hat{U}^\dagger$ cannot be fully disentangled.
Nevertheless, the small parameter corresponding to the ratio of the electron-phonon coupling to the phonon frequency $g_q/\omega$ enables a low-order expansion  in $g_q/\omega$ in the transformed frame $\hat{U}$ and, following a polaron-hopping disentanglement, we can approximate $\hat{H}' \approx \hat{H}_\mathrm{eff}$, where
\begin{equation}
    \begin{aligned}
        \hat{H}_\text{eff}
         =& -J_\text{eff}\sum_{i=1}^L \sum_{\sigma = \uparrow, \downarrow} \left(
                \hat{c}_{i,\sigma}^\dagger \hat{c}_{i+1,\sigma}^\nodagger + \text{h.c.}
            \right ) \\
        &+U_\text{eff} \sum_{i=1}^L \left( \hat{n}_{i,\uparrow} -\frac{1}{2} \right )\left( \hat{n}_{i,\downarrow} -\frac{1}{2} \right ) \left(
                \hat{\beta}_i^\dagger \hat{\beta}_i^\nodagger + \frac{1}{2}
            \right ) \\
        & + \epsilon_\text{eff} \sum_{i=1}^L  \left( \hat{n}_i - 1 \right ) \left(
                \hat{\beta}_i^\dagger \hat{\beta}_i^\nodagger + \frac{1}{2}
            \right ) +\omega_\text{eff} \sum_{i=1}^L
            \left(
                \hat{\beta}_i^\dagger \hat{\beta}_i^\nodagger + \frac{1}{2}
            \right )
        \; .
    \end{aligned}
    \label{eq:effective:hamiltonian}
    \end{equation}
The first term in~\cref{eq:effective:hamiltonian} describes electron hopping with a renormalized amplitude $J_\text{eff} = J e^{-\frac{1}{2}(\frac{g_q}{\omega})^2(\alpha^4 + 2\alpha^2+1)}$, the second describes a pump-induced Hubbard attraction with $U_\text{eff} = -4g_q^2/2\omega$, the third represents a pump-induced onsite electron potential with $\epsilon_\text{eff} = 2g_q$ and in the last term the phonon frequency has been modified to $\omega_\text{eff} = \omega-g_q^2/\omega$.
An important property of~\cref{eq:effective:hamiltonian} is that the electrons conserve the phonon dynamics: The terms $\hat{\beta}_i^\dagger \hat{\beta}_i^\nodagger$ simply count the number of squeezed phonon excitations on site $i$. 
We keep $g_q/\omega$ fixed and consider three phonon frequencies corresponding to fast $(\omega=\pi/2)J$, intermediate $(\omega=\pi/5)J$, and slow $(\omega=\pi/10)J$ phonons. All the corresponding effective parameters are listed in Table~\ref{tab:eff:model}.  
Figure 4 of the main text compares momentum-resolved charge correlations and double occupancy from the effective model with the full simulation of the exact model, while Figures~\ref{fig:supp:pairing:effective} and~\ref{fig:supp:spin:effective} here show analogous results for pairing $P_k(t)$ and spin $S_k(t)$, demonstrating that the effective Hamiltonian captures the qualitative features of the exact model.

\begin{table}[h!]
\centering
\begin{tabular}{|>{\columncolor[gray]{0.9}}c|c|c|c|c|}
\hline
\rowcolor[gray]{0.9} % Highlight the first row
$g_q$, $\mathbf{\omega}$ & $U_\mathrm{eff} = \frac{-4g_q^2}{\omega}$ & $\epsilon_\mathrm{eff} = 2g_q$ & $\omega_\mathrm{eff} = \omega-g_q^2/\omega$ & $J_\mathrm{eff} = J e^{-\frac{1}{2}(\frac{g_q}{\omega})^2(\alpha^4 + 2\alpha^2+1)}$  \\ \hline
$0.35$, $\pi/2$ & -0.312 & 0.70 & 1.493 & 0.799 \\ \hline
$0.14$, $\pi/5$ & -0.125 & 0.28 & 0.597 & 0.799 \\ \hline
$0.07$, $\pi/10$ & -0.062 & 0.14 & 0.299 & 0.799 \\ \hline
\end{tabular}
\caption{Parameters of the effective model \cref{eq:effective:hamiltonian} for the three values of $(g_q,\omega)$ used in this work, with fixed ratio $g_q/\omega \approx 0.22$,  and $\alpha=\sqrt{2}$. All quantities are  in units of the hopping $J$.}
\label{tab:eff:model}
\end{table}


\section{\label{app:sec:pst} Phonon state tomography}

In this section, we briefly summarize the~\acrfull{PST} scheme~\cite{moroder2024}, used in the main text to analyze how lowering the phonon frequency~$\omega$ mitigates phonon-induced disorder.  
In essence, \gls{PST} reveals how phonon excitations shape electronic observables.

Let $\mathbf n_{\mathrm{ph}} = (n_0,n_1,\ldots,n_{L_{\mathrm{ph}}-1})$ denote the phonon occupation on each site (cf.\ Figure~\ref{fig:app:schematic:pp}).  
We rewrite the electron–phonon wavefunction as  
\begin{equation}
\ket{\Psi} \;=\;
   \sum_{\{\mathbf n_{\mathrm{ph}}\}}
   \varphi(\mathbf n_{\mathrm{ph}})
   \ket{\psi_{\mathrm{el}}(\mathbf n_{\mathrm{ph}})}
   \otimes
   \ket{\mathbf n_{\mathrm{ph}}}, 
\label{eq:psi:for:pst}
\end{equation}
where $\ket{\psi_{\mathrm{el}}(\mathbf n_{\mathrm{ph}})}\in\mathcal H_{\mathrm{el}}$ is a normalized electronic state and 
$|\varphi(\mathbf n_{\mathrm{ph}})|^{2}\equiv p(\mathbf n_{\mathrm{ph}})$ is the probability of the phonon configuration~$\mathbf n_{\mathrm{ph}}$.

For any electronic operator~$\hat A$, the expectation value splits exactly into  
\begin{equation}
\langle\hat A\rangle
   =\sum_{\{\mathbf n_{\mathrm{ph}}\}}
     p(\mathbf n_{\mathrm{ph}})
     \bra{\psi_{\mathrm{el}}(\mathbf n_{\mathrm{ph}})}\hat A
     \ket{\psi_{\mathrm{el}}(\mathbf n_{\mathrm{ph}})}
   \;\equiv\;
   \sum_{\{\mathbf n_{\mathrm{ph}}\}}
     p(\mathbf n_{\mathrm{ph}})\,A(\mathbf n_{\mathrm{ph}}).
\label{eq:app:pst:exact:expectation}
\end{equation}
Because the number of configurations grows exponentially, evaluating every $A(\mathbf n_{\mathrm{ph}})$ is impractical.  
Instead, we sample a manageable set $\Omega$ according to~$p(\mathbf n_{\mathrm{ph}})$ and approximate  
\begin{equation}
\langle\hat A\rangle
   \approx
   \frac{1}{|\Omega|}
   \sum_{\mathbf n_{\mathrm{ph}}\in\Omega}
   \bra{\psi_{\mathrm{el}}(\mathbf n_{\mathrm{ph}})}\hat A
   \ket{\psi_{\mathrm{el}}(\mathbf n_{\mathrm{ph}})}.
\label{eq:app:pst:approx:expectation}
\end{equation}
Grouping configurations by the total phonon number 
$N_{\mathrm{ph}}=\sum_i n_i$ yields both 
$A(N_{\mathrm{ph}})$ and the global distribution 
$\mathcal P(N_{\mathrm{ph}})$.

The unknown global law $p(\mathbf n_{\mathrm{ph}})$ is factorized into conditional probabilities\,\cite{Ferris2012},  
\begin{equation}
p(\mathbf n_{\mathrm{ph}})
   = p(n_0)\,
     p(n_1|n_0)\,\cdots\,
     p(n_{L-1} | n_{L-2},\ldots,n_0),
\label{eq:app:pst:factorization}
\end{equation}
allowing one to generate a sample by sweeping across the chain and performing successive projective measurements, accessed through the phonon~\gls{1RDM}.  
Within the projected–purified \gls{MPS} representation, this reduces to efficient \glspl{SVD}\,\cite{Koehler2021} rather than expensive density-matrix evaluations.

\begin{figure}[H]
    \centering
    \includegraphics[width=0.8\linewidth]{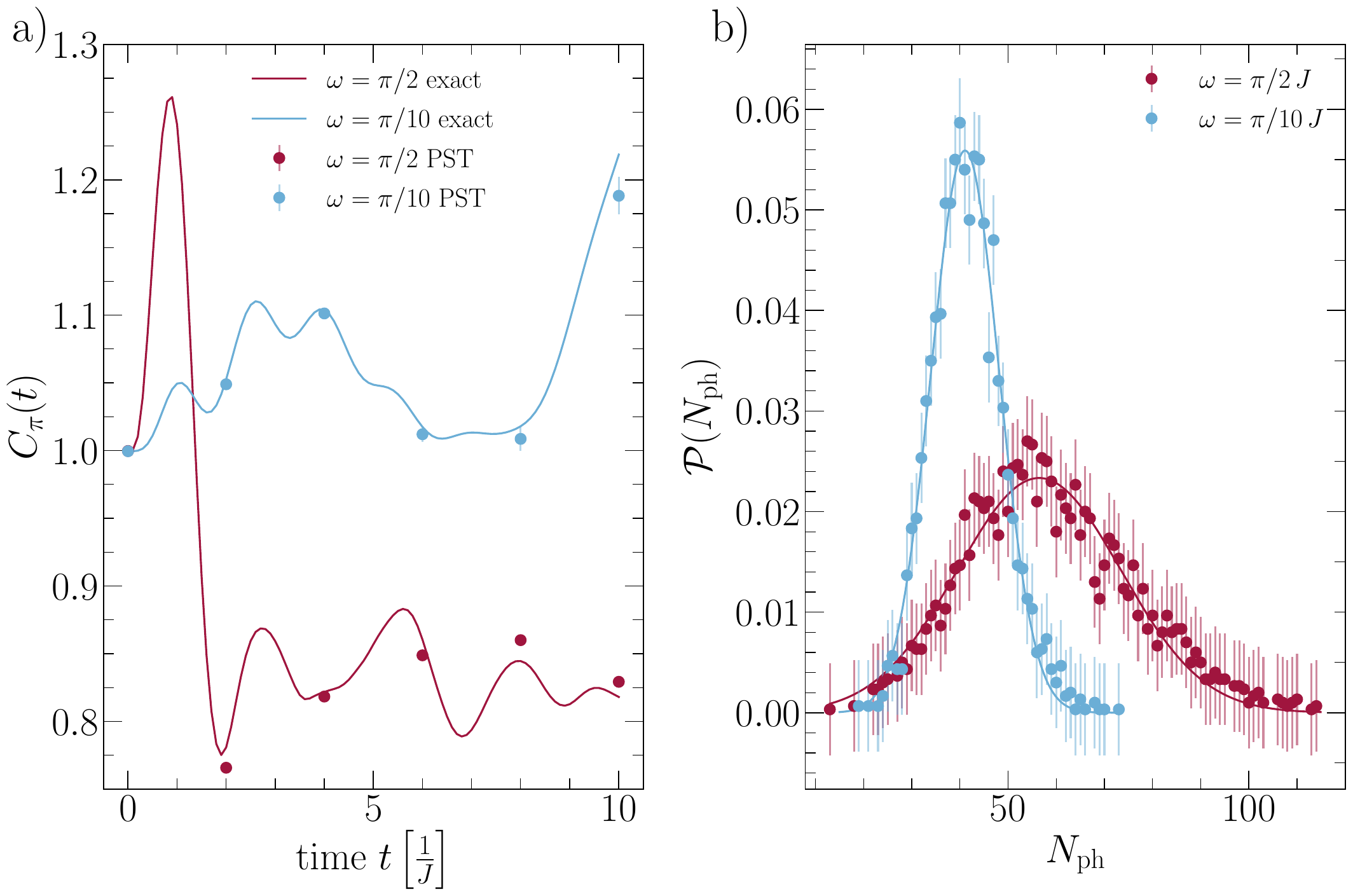}
    \caption{
    Panel a): Time evolution of the $k=\pi$ charge correlator, comparing PST estimates (symbols) with the exact results (solid curves).  Panel b): Global phonon‐number distribution $\mathcal{P}(N_{\mathrm{ph}})$ obtained from the same PST sampling procedure.  We employ $|\Omega| = 3000$ samples on a chain of $L = 20$ sites, with a phonon cutoff $d_\nu^{\max} = 40$ and a maximum bond dimension $\chi_{\max} = 2000$.  In panel a) the error bars represent the standard error $\epsilon = \sigma / \sqrt{|\Omega|}$. Because $\mathcal{P}(N_{\mathrm{ph}})$ is intrinsically noisier, its error bars are displayed at $\epsilon/4$ to keep the figure readable.
    }
    \label{fig:app:pst}
\end{figure}

Figure 5 of the main text employs \gls{PST} to study the late-time behavior ($tJ=10$) of a one-dimensional photoexcited metal, analyzing both the momentum-resolved charge correlator $C_k$ (defined in Figure~\ref{fig:app:Ck:L20:L32}) and the phonon distribution $\mathcal P(N_{\mathrm{ph}})$.  
To assess accuracy, Figure~\ref{fig:app:pst}(a) compares $\langle C_\pi(t)\rangle_{\mathrm{PST}}$ obtained from~\eqref{eq:app:pst:approx:expectation} with the exact value $\langle\Psi(t)|\hat C_\pi|\Psi(t)\rangle$.  
The \gls{PST} estimates (circles) improve as the phonon frequency decreases, because for large~$\omega$ the late-time distribution $\mathcal P(N_{\mathrm{ph}})$, shown in Figure~\ref{fig:app:pst}(b), spreads over many more $N_{\mathrm{ph}}$ sectors, making sampling harder.

\section{Additional results}

\begin{figure}[H]
    \centering
    \includegraphics[width=0.6\linewidth]{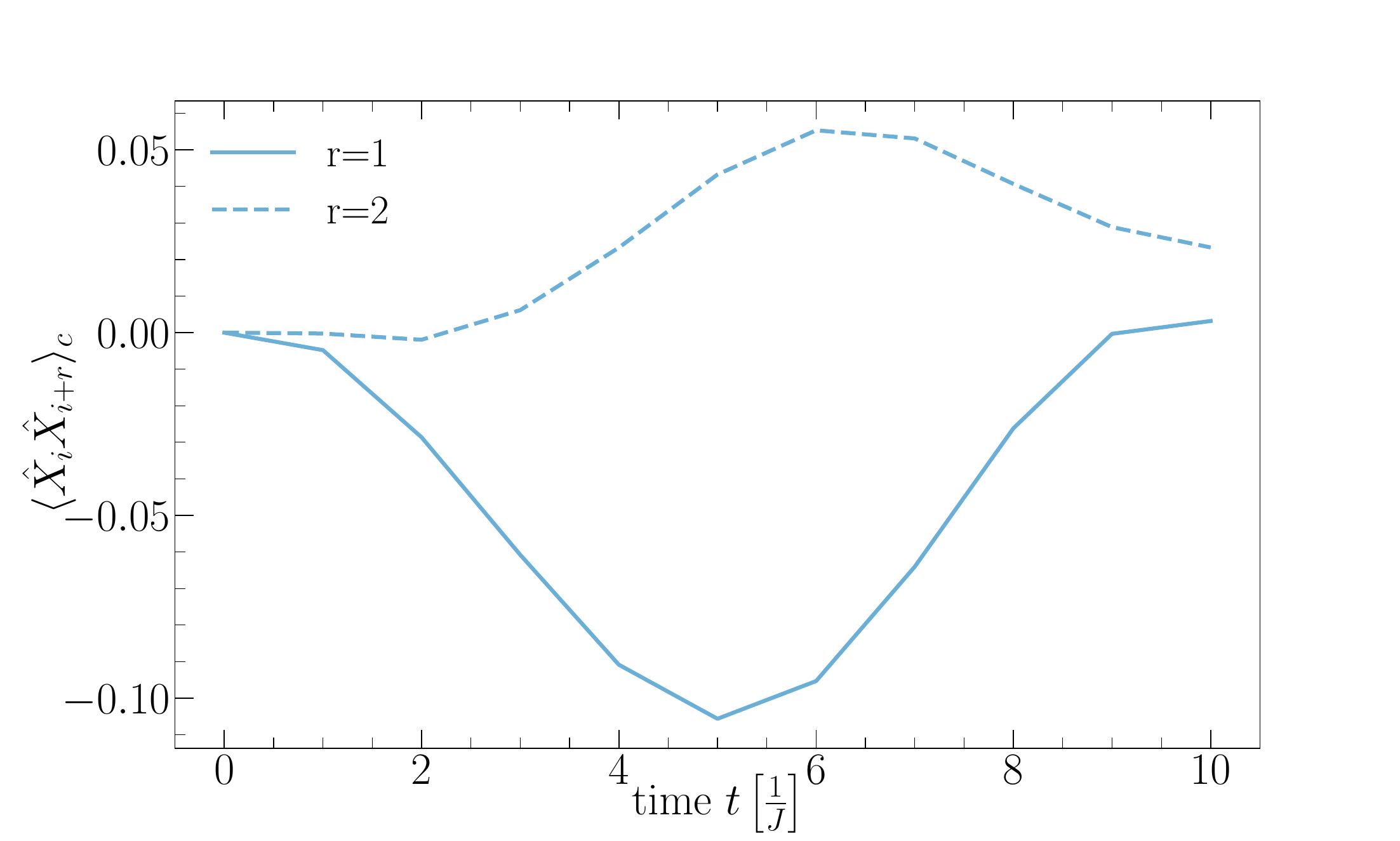}
    \caption{Connected phonon–displacement correlator   $\langle \hat{X}_i \hat{X}_{i+r}\rangle_c
      \equiv
      \langle \hat{X}_i \hat{X}_{i+r}\rangle
      -\langle \hat{X}_i\rangle\langle\hat{X}_{i+r}\rangle$   with $\hat{X}_i=(\hat{b}_i^\dagger+\hat{b}_i)/\sqrt{2}$.  Nearest neighbors exhibit anticorrelations, while next-nearest neighbors show positive correlations. Here, the phonon frequency $\omega=\pi/10\,J$, $g_q =0.07 \,J$, $\alpha = \sqrt{2}$, with chain length $L=20$, maximum bond dimension $\chi_{\max}=2000$, and phonon cutoff $d_\nu^{\max}=40$.
}
    \label{fig:enter-label}
\end{figure}

\begin{figure}[!t]
    \centering
    \includegraphics[width=0.7\linewidth]{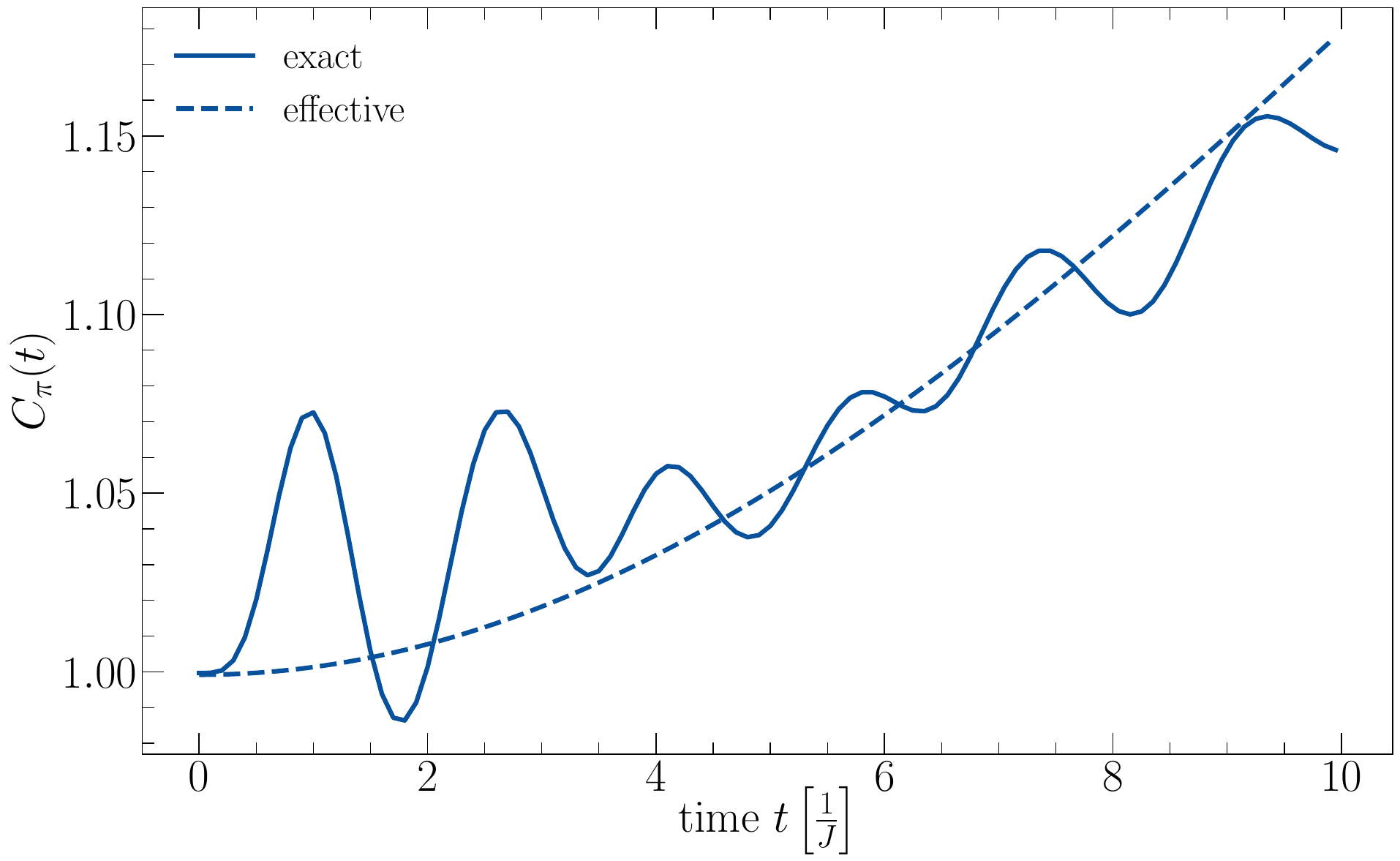}
a    \caption{The zero phonon frequency limit. We show the dynamics of the charge correlation function at momentum $k=\pi$ for the exact model (solid line) and the effective model (dashed line). We set $L=20$, $g_q=0.1\,J$, $\alpha = \sqrt{2}$, $\chi_{\max}=2000$, and $d_\nu^{\max}=40$. 
For the exact model we take $\omega=0\,J$, whereas for the effective model we use $\omega=0.01 \,J$.
}
\label{fig:app:zero:frequency}
\end{figure}

\begin{figure}
    \centering
    \includegraphics[width=0.9\linewidth]{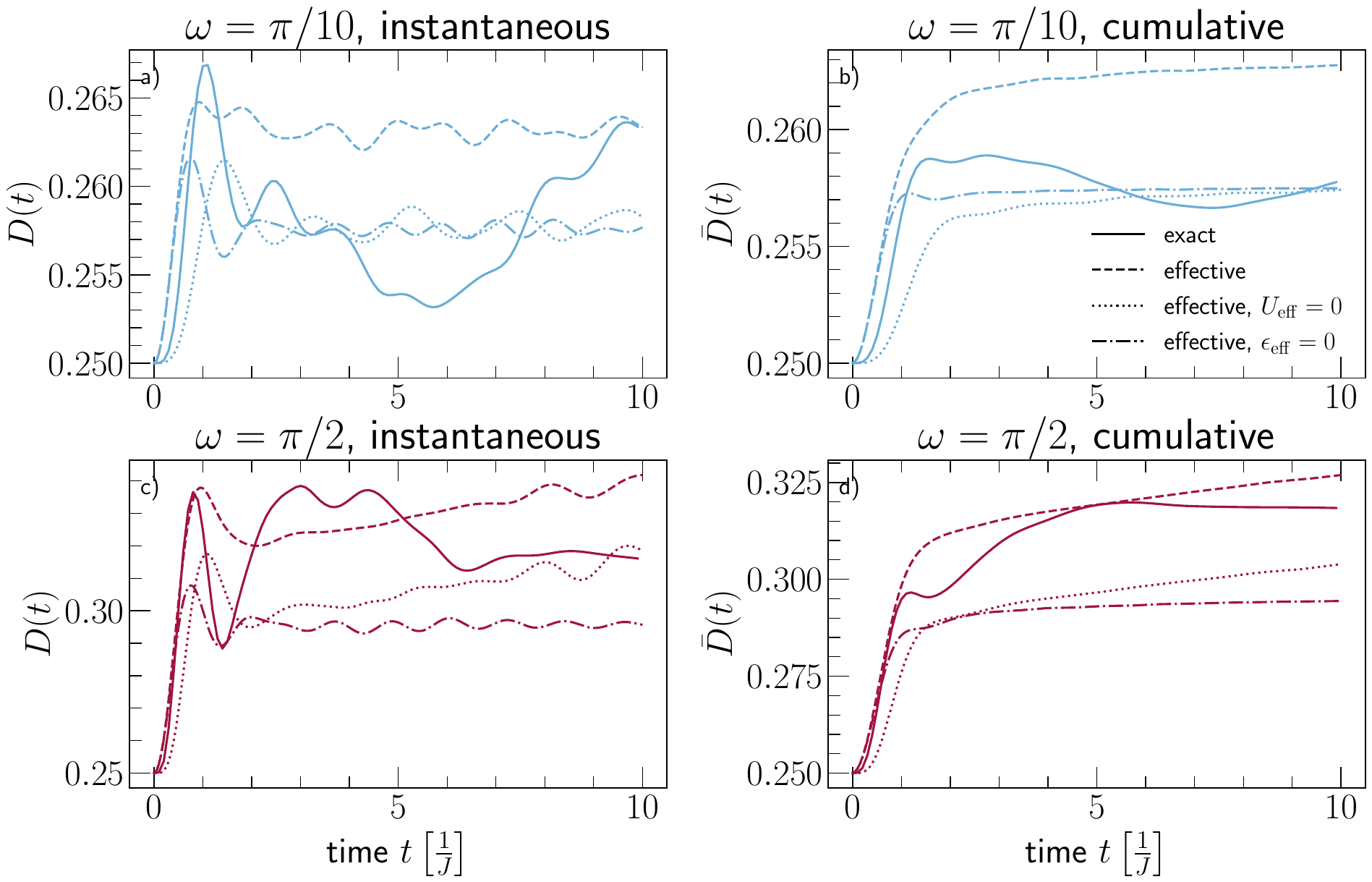}
    \caption{ We compare the exact dynamics of the double occupancy $D(t)\equiv \frac{1}{L}\sum_i \langle \Psi(t)| \hat n_{i\uparrow}\hat n_{i\downarrow} |\Psi(t)\rangle$ generated by \cref{app:eq:total:hamiltonian} (solid) with those from (i) the full effective Hamiltonian \cref{eq:effective:hamiltonian} (dashed), (ii) the effective Hamiltonian without interactions ($U_{\mathrm{eff}}=0$, dotted), and (iii) the effective Hamiltonian without disorder ($\epsilon_{\mathrm{eff}}=0$, dash-dotted). The left column shows the instantaneous $D(t)$, while the right column displays the time-averaged quantity $\bar D(t)=\frac{1}{t}\int_0^t D(t)\,dt$.  We fix $g_q/\omega = 0.7/\pi \approx 0.22$, $\alpha = \sqrt{2}$, and set $L=20$, $\chi_{\max}=2000$, and $d_\nu^{\max}=40$.
}
    \label{fig:double:occ:eff}
\end{figure}

\begin{figure}
    \centering
    \includegraphics[width=0.9\linewidth]{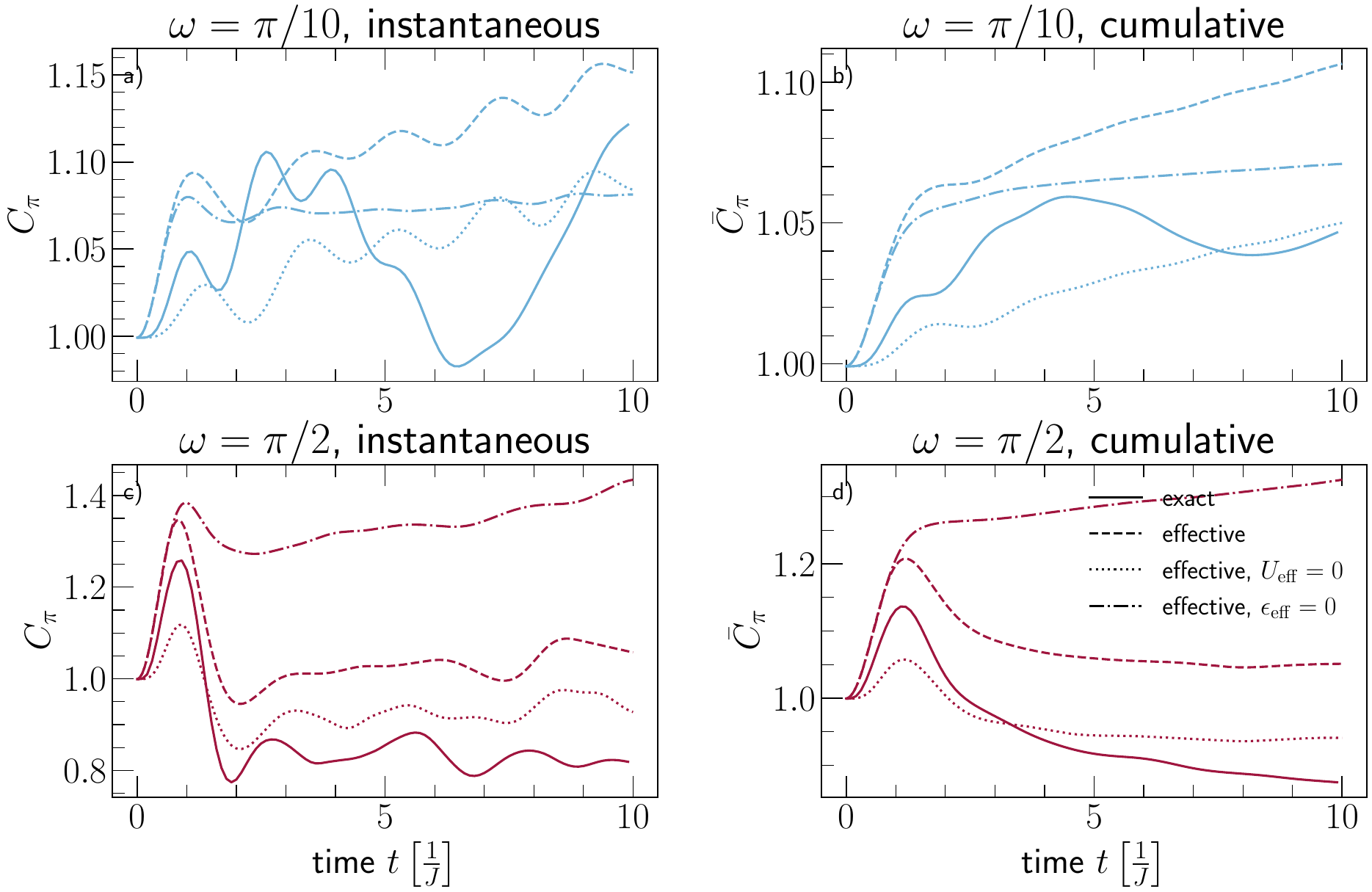}
    \caption{We compare the exact dynamics of the momentum-resolved charge correlator $C_k(t)$ at $k=\pi$ generated by \cref{app:eq:total:hamiltonian} (solid) with those from (i) the full effective Hamiltonian \cref{eq:effective:hamiltonian} (dashed), (ii) the effective Hamiltonian with the interaction removed ($U_{\mathrm{eff}}=0$, dotted), and (iii) the effective Hamiltonian without disorder ($\epsilon_{\mathrm{eff}}=0$, dash-dotted).   The left column shows the instantaneous $C_\pi(t)$, while the right column displays the time-averaged quantity $\bar{C}_\pi(t)=\frac{1}{t}\!\int_0^t C_\pi(t)\,dt$.   We fix $g_q/\omega = 0.7/\pi \approx 0.22$, $\alpha = \sqrt{2}$,  and set $L=20$, $\chi_{\max}=2000$, and $d_\nu^{\max}=40$.
}
    \label{fig:C:pi:eff}
\end{figure}

\begin{figure}
    \centering
    \includegraphics[width=0.9\linewidth]{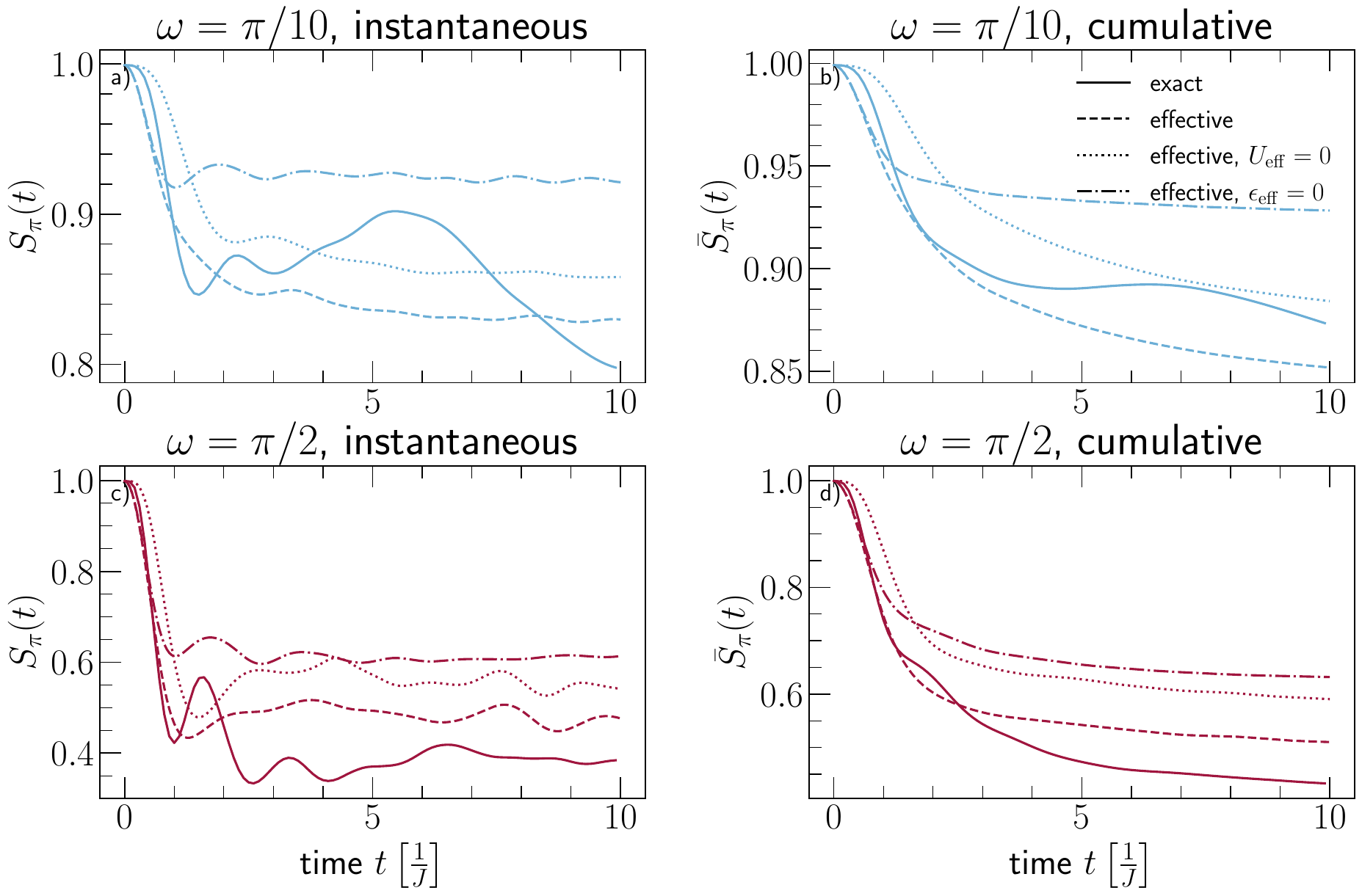}
    \caption{We compare the exact dynamics of the momentum-resolved spin correlator $S_k(t)$ at $k=\pi$ (solid) with those from (i) the full effective Hamiltonian \cref{eq:effective:hamiltonian} (dashed), (ii) the effective Hamiltonian without interactions ($U_{\mathrm{eff}}=0$, dotted), and (iii) the effective Hamiltonian without disorder ($\epsilon_{\mathrm{eff}}=0$, dash-dotted).  The left column shows the instantaneous $S_\pi(t)$, while the right column displays the time-averaged quantity $\bar S_\pi(t)=\frac{1}{t}\int_0^t S_\pi(t)\,dt$.  We fix $g_q/\omega = 0.7/\pi \approx 0.22$, $\alpha = \sqrt{2}$, and set $L=20$, $\chi_{\max}=2000$, and $d_\nu^{\max}=40$.
}
    \label{fig:S:pi:eff}
\end{figure}

Supplementary results are collected in 
Figs.~\ref{fig:enter-label}, 
\ref{fig:app:zero:frequency}, 
\ref{fig:double:occ:eff}, 
\ref{fig:C:pi:eff}, and 
\ref{fig:S:pi:eff}.

\paragraph*{Phonon correlator.}
Figure~\ref{fig:enter-label} shows the connected phonon–displacement correlation
$\langle \hat{X}_i \hat{X}_{i+r}\rangle_c
      \equiv
      \langle \hat{X}_i \hat{X}_{i+r}\rangle
      -\langle \hat{X}_i\rangle\langle\hat{X}_{i+r}\rangle$,
with
$\hat{X}_i = (\hat{b}_i^\dagger + \hat{b}_i)/\sqrt{2}$ 
(as sketched in Fig.~1(d) of the main text).  
Nearest neighbors become negatively correlated, whereas next-nearest neighbors become positively correlated, signaling enhanced staggered lattice displacements.

\paragraph*{Enhanced coherence from slow phonons.}
Figure~\ref{fig:app:zero:frequency} examines the $\omega\to0$ limit—nominally unstable in the nonlinear model but nonetheless informative to clarify the role of slow phonons.  
Staggered charge correlations grow dramatically, and the agreement between the effective and exact simulations improves further.

\paragraph*{Interplay of pump-induced disorder and electron attraction in the effective model.}
Figures~\ref{fig:double:occ:eff}, \ref{fig:C:pi:eff}, and \ref{fig:S:pi:eff} dissect the roles of onsite disorder and the effective Hubbard attraction.  The qualitative trends largely persist even for $U_{\mathrm{eff}}=0$, indicating that disorder by itself governs much of the time-averaged correlation dynamics of the full model.  Discrepancies—arising from the subtle interplay between interaction and disorder—remain and will be investigated in future work.

\FloatBarrier
\bibliography{Literature}